\documentclass[useAMS,usenatbib,onecolumn,usegraphicx]{mn2e}

\usepackage{bm}

\def\be{\begin{equation}}
\def\ee{\end{equation}}
\def\ba{\begin{eqnarray}}
\def\ea{\end{eqnarray}}

\title[Magnetic neutron star surfaces and polar caps]
{Condensed surfaces of magnetic neutron stars, thermal
surface emission, and particle acceleration above pulsar polar caps}

\author[Zach Medin and Dong Lai]{Zach Medin\thanks{Email:
zach@astro.cornell.edu}
and Dong Lai\thanks{Email: dong@astro.cornell.edu}\\
Department of Astronomy, Center for Radiophysics and Space Research,
Cornell University, Ithaca, NY 14853, USA}

\begin{document}

\date{Accepted 2007 September 18. Received 2007 September 17; in original form 2007 August 28}

\pagerange{\pageref{firstpage}--\pageref{lastpage}} \pubyear{2007}

\label{firstpage}

\maketitle

\begin{abstract}
Recent calculations indicate that the cohesive energy of condensed
matter increases with magnetic field strength and becomes very
significant at magnetar-like fields (e.g., 10~keV at $3\times
10^{14}$~G for zero-pressure condensed iron). This implies that for
sufficiently strong magnetic fields and/or low temperatures, the
neutron star surface may be in a condensed state with little gas or
plasma above it. Such surface condensation can significantly affect
the thermal emission from isolated neutron stars, and may lead to the
formation of a charge-depleted acceleration zone (``vacuum gap'') in
the magnetosphere above the stellar polar cap. Using the latest
results on the cohesive property of magnetic condensed matter, we
quantitatively determine the conditions for surface condensation and
vacuum gap formation in magnetic neutron stars. We find that
condensation can occur if the thermal energy $kT$ of the neutron star
surface is less than about 8\% of its cohesive energy $Q_s$, and that
a vacuum gap can form if $\mathbf{\Omega} \cdot \mathbf{B_p}<0$ (i.e.,
the neutron star's rotation axis and magnetic moment point in opposite
directions) and $kT$ is less than about 4\% of $Q_s$. For example, at
$B=3\times 10^{14}$~G, a condensed Fe surface forms when $T\la 10^7$~K
and a vacuum gap forms when $T\la 5\times 10^6$~K.  Thus, vacuum gap
accelerators may exist for some neutron stars. Motivated by this
result, we also study the physics of pair cascades in the
(Ruderman-Sutherland type) vacuum gap model for photon emission by
accelerating electrons and positrons due to both curvature radiation
and resonant/nonresonant inverse Compton scattering. Our calculations
of the condition of cascade-induced vacuum breakdown and the related
pulsar death line/boundary generalize previous works to the
superstrong field regime. We find that inverse Compton scatterings do
not produce a sufficient number of high energy photons in the gap
(despite the fact that resonantly upscattered photons can immediately
produce pairs for $B\ga 1.6\times 10^{14}$~G) and thus do not lead to
pair cascades for most neutron star parameters (spin and magnetic
field). We discuss the implications of our results for the recent
observations of neutron star thermal radiation as well as for the
detection/non-detection of radio emission from high-B pulsars and
magnetars.
\end{abstract}


\section{Introduction}
\label{sec:intro}

Recent observations of neutron stars have provided a wealth of
information on these objects, but they have also raised many new
questions. For example, with the advent of X-ray telescopes such as
Chandra and XMM-Newton, detailed observations of the thermal radiation
from the neutron star surface have become possible. These observations
show that some nearby isolated neutron stars (e.g., RX J1856.5-3754)
appear to have featureless, nearly blackbody spectra
(\citealt{burwitz03,vankerkwijk07}). Radiation from a bare condensed
surface (where the overlying atmosphere has negligible optical depth)
has been invoked to explain this nearly perfect blackbody emission
(e.g.,
\citealt{burwitz03,mori03,turolla04,vanadelsberg05,perez06,ho07}; but
see \citealt{ruderman03} for an alternative view). However, whether
surface condensation actually occurs depends on the cohesive
properties of the surface matter (e.g., \citealt{lai01}).

Equally puzzling are the observations of anomalous X-ray pulsars
(AXPs) and soft gamma-ray repeaters (SGRs) (see \citealt{woods05} for
a review). Though these stars are believed to be magnetars, neutron
stars with extremely strong magnetic fields ($B \ga 10^{14}$~G), they
mostly show no pulsed radio emission (but see
\citealt{camilo06,camilo07,kramer07}) and their X-ray radiation is too
strong to be powered by rotational energy loss. By contrast, several
high-B radio pulsars with inferred surface field strengths similar to
those of magnetars have been discovered (e.g.,
\citealt*{mclaugh05,vranevsevic07}). A deeper understanding of the
distinction between pulsars and magnetars requires further
investigation of the mechanisms by which pulsars and magnetars radiate
and of their magnetospheres where this emission
originates. Theoretical models of pulsar and magnetar magnetospheres
depend on the cohesive properties of the surface matter in strong
magnetic fields (e.g.,
\citealt*{ruderman75,arons79,cheng80,usov96,harding98,gil03,muslimov03,beloborodov07}). For
example, depending on how strongly bound the surface matter is, a
charge-depleted acceleration zone (``vacuum gap'') above the polar cap
of a pulsar may or may not form, and this will affect pulsar radio
emission and other high-energy emission processes.

The cohesive property of the neutron star surface matter plays a key
role in these and other neutron star processes and observed
phenomena. The cohesive energy refers to the energy required to pull
an atom out of the bulk condensed matter at zero pressure. A related
(but distinct) quantity is the electron work function, the energy
required to pull out an electron. For magnetized neutron star surfaces
the cohesive energy and work function can be many times the
corresponding terrestrial values, due to the strong magnetic fields
threading the matter (e.g., \citealt{ruderman74,lai01}).

In two recent papers (\citealt{medin06a,medin06b}, hereafter ML06a,b),
we carried out detailed, first-principle calculations of the cohesive
properties of H, He, C, and Fe surfaces at field strengths between $B
= 10^{12}$~G to $2\times10^{15}$~G. The main purpose of this paper is
to investigate several important astrophysical implications of these
results (some preliminary investigations were reported in
\citealt{medin07}). This paper is organized as follows. In
Section~\ref{sect:cohes} we briefly summarize the key results
(cohesive energy and work function values) of ML06a,b used in this
paper. In Section~\ref{sect:cond} we examine the possible formation of
a bare neutron star surface, which directly affects the surface
thermal emission. We find that the critical temperature below which a
phase transition to the condensed state occurs is approximately given
by $kT_{\rm crit} \sim 0.08Q_s$, where $Q_s$ is the cohesive energy of
the surface. In Section~\ref{sect:polar} we consider the conditions
for the formation of a polar vacuum gap in pulsars and magnetars. We
find that neutron stars with rotation axis and magnetic moment given
by $\mathbf{\Omega} \cdot \mathbf{B_p}>0$ are unable to form vacuum
gaps (since the electrons which are required to fill the gaps can be
easily supplied by the surface), but neutron stars with
$\mathbf{\Omega} \cdot \mathbf{B_p}<0$ can form vacuum gaps provided
that the surface temperature is less than $kT_{\rm crit} \sim 0.04Q_s$
(and that particle bombardment does not completely destroy the gap;
see Section~\ref{sect:discuss}). In Section~\ref{sect:death} we
discuss polar gap radiation mechanisms and the pulsar death
line/boundary in the vacuum gap model. We find that when curvature
radiation is the dominant radiation mechanism in the gap, a pair
cascade is possible for a large range of parameter space (in the
$P$--$\dot{P}$ diagram), but when inverse Compton scattering (either
resonant or nonresonant) is the dominant radiation mechansim, vacuum
breakdown is possible for only a very small range of parameter
values. Implications of our results for recent observations are
discussed in Section~\ref{sect:discuss}. Some technical details (on
our treatment of inverse Compton scattering and vacuum gap
electrodynamics of oblique rotators) are given in two appendices.

\section{Cohesive Properties of Condensed Matter in Strong Magnetic Fields}
\label{sect:cohes}

It is well-known that the properties of matter can be drastically
modified by strong magnetic fields. The natural atomic unit for the
magnetic field strength, $B_0$, is set by equating the electron
cyclotron energy $\hbar\omega_{ce}=\hbar
(eB/m_ec)=11.577\,B_{12}$~keV, where $B_{12}=B/(10^{12}~{\rm G})$, to
the characteristic atomic energy $e^2/a_0=2\times 13.6$~eV (where
$a_0$ is the Bohr radius):
\be
B_0=\frac{m_e^2e^3c}{\hbar^3}=2.3505\times 10^9\, {\rm G}.
\label{eqb0}
\ee
For $b=B/B_0\ga 1$, the usual perturbative treatment of the magnetic 
effects on matter (e.g., Zeeman splitting of atomic
energy levels) does not apply. Instead, the Coulomb forces act as a
perturbation to the magnetic forces, and the electrons in an atom
settle into the ground Landau level. Because of the extreme
confinement of the electrons in the transverse
direction (perpendicular to the field), the Coulomb force becomes much
more effective in binding the electrons along the magnetic field
direction. The atom attains a cylindrical structure. Moreover, it is
possible for these elongated atoms to form molecular chains by
covalent bonding along the field direction. Interactions between the
linear chains can then lead to the formation of three-dimensional
condensed matter (\citealt{ruderman74,ruder94,lai01}).

The basic properties of magnetized condensed matter can be estimated
using the uniform electron gas model (e.g., \citealt{kadom70}). The
energy per cell of a zero-pressure condensed matter is given by
\be
{\cal E}_s\sim -120\,Z^{9/5}B_{12}^{2/5}~{\rm eV},
\ee
and the corresponding condensation density is 
\be
\rho_{s}\sim 560\,A\,Z^{-3/5}B_{12}^{6/5}\,{\rm g~cm}^{-3},
\ee
where $Z,~A$ are the charge number and mass number of the ion (see
\citealt{lai01} and references therein for further refinements to the
uniform gas model). Although this simple model gives a reasonable
estimate of the binding energy for the condensed state, it is not
adequate for determining the cohesive property of the condensed
matter. The cohesive energy is the (relatively small) difference
between the atomic ground-state energy ${\cal E}_a$ and the zero-pressure
condensed matter energy ${\cal E}_s$, both increasing rapidly with
$B$. Moreover, the electron Fermi energy (including both kinetic
energy and Coulomb energy) in the uniform gas model,
\be
\varepsilon_F=(3/5Z){\cal E}_s\sim -73Z^{4/5}B_{12}^{2/5}~{\rm eV},
\label{fermieq}
\ee
may not give
a good scaling relation for the electron work function when detailed
electron energy levels (bands) in the condensed matter are taken into
account.

There have been few quantitative studies of infinite chains and
zero-pressure condensed matter in strong magnetic fields. Earlier
variational calculations (e.g., \citealt{flowers77,muller84}) as well
as calculations based on Thomas-Fermi type statistical models
(e.g., \citealt{abrahams91,fushiki92}), while useful in establishing
scaling relations and providing approximate energies of the atoms and
the condensed matter, are not adequate for obtaining reliable energy
differences (cohesive energies). Quantitative results for the energies
of infinite chains of hydrogen molecules H$_\infty$ over a wide range of
field strengths ($B\gg B_0$) were presented in \citet{lai92} (using
the Hartree-Fock method with the plane-wave approximation; see also
\citealt{lai01} for some results for He$_\infty$) and in
\citet{relovsky96} (using density functional theory). For heavier
elements such as C and Fe, the cohesive energies of one dimensional
(1D) chains have only been calculated at a few magnetic field
strengths in the range of $B=10^{12}$--$10^{13}$~G, using Hartree-Fock
models \citep{neuhauser87} and density functional theory
\citep{jones85}. There were some discrepancies between the results of
these works, and some adopted a crude treatment for the band structure
\citep{neuhauser87}. An approximate calculation of 3D condensed matter
based on density functional theory was presented in \citet{jones86}.

Our calculations of atoms and small molecules (ML06a) and of infinite
chains and condensed matter (ML06b) are based on a newly developed
density functional theory code. Although the Hartree-Fock method is
expected to be highly accurate in the strong field regime, it becomes
increasingly impractical for many-electron systems as the magnetic
field increases, since more and more Landau orbitals are occupied
(even though electrons remain in the ground Landau level) and keeping
track of the direct and exchange interactions between electrons in
various orbitals becomes computational rather tedious. Compared to
previous density-functional theory calculations, we used an improved
exchange-correlation function for highly magnetized electron gases,
and we calibrated our density-functional code with previous results
(when available) based on other methods. Most importantly, in our
calculations of 1D condensed matter, we treated the band structure of
electrons in different Landau orbitals self-consistently without
adopting ad-hoc simplifications. This is important for obtaining
reliable results for the condensed matter. Since each Landau orbital
has its own energy band, the number of bands that need to be
calculated increases with $Z$ and $B$, making the computation
increasingly complex for superstrong magnetic field strengths
(e.g., the number of occupied bands for Fe chains at $B=2\times
10^{15}$~G reaches 155; see Fig.~16 of ML06b). Our density-functional
calculations allow us to obtain the energies of atoms and small
molecules and the energy of condensed matter using the same method,
thus providing reliable cohesive energy and work function values for
condensed surfaces of magnetic neutron stars.

In ML06a, we described our calculations for various atoms and
molecules in magnetic fields ranging from $10^{12}$~G to $2\times
10^{15}$~G for H, He, C, and Fe, representative of the most likely
neutron star surface compositions. Numerical results of the
ground-state energies are given for H$_N$ (up to $N=10$), He$_N$ (up
to $N=8$), C$_N$ (up to $N=5$), and Fe$_N$ (up to $N=3$), as well as
for various ionized atoms. In ML06b, we described our calculations for
infinite chains for H, He, C, and Fe in that same magnetic field
range. For relatively low field strengths, chain-chain interactions
play an important role in the cohesion of three-dimensional (3D)
condensed matter. An approximate calculation of 3D condensed matter is
also presented in ML06b. Numerical results of the ground-state and
cohesive energies, as well as the electron work function and the
zero-pressure condensed matter density, are given in ML06b for
H$_\infty$ and H(3D), He$_\infty$ and He(3D), C$_\infty$ and C(3D),
and Fe$_\infty$ and Fe(3D).

\begin{figure}
\includegraphics[width=6.5in]{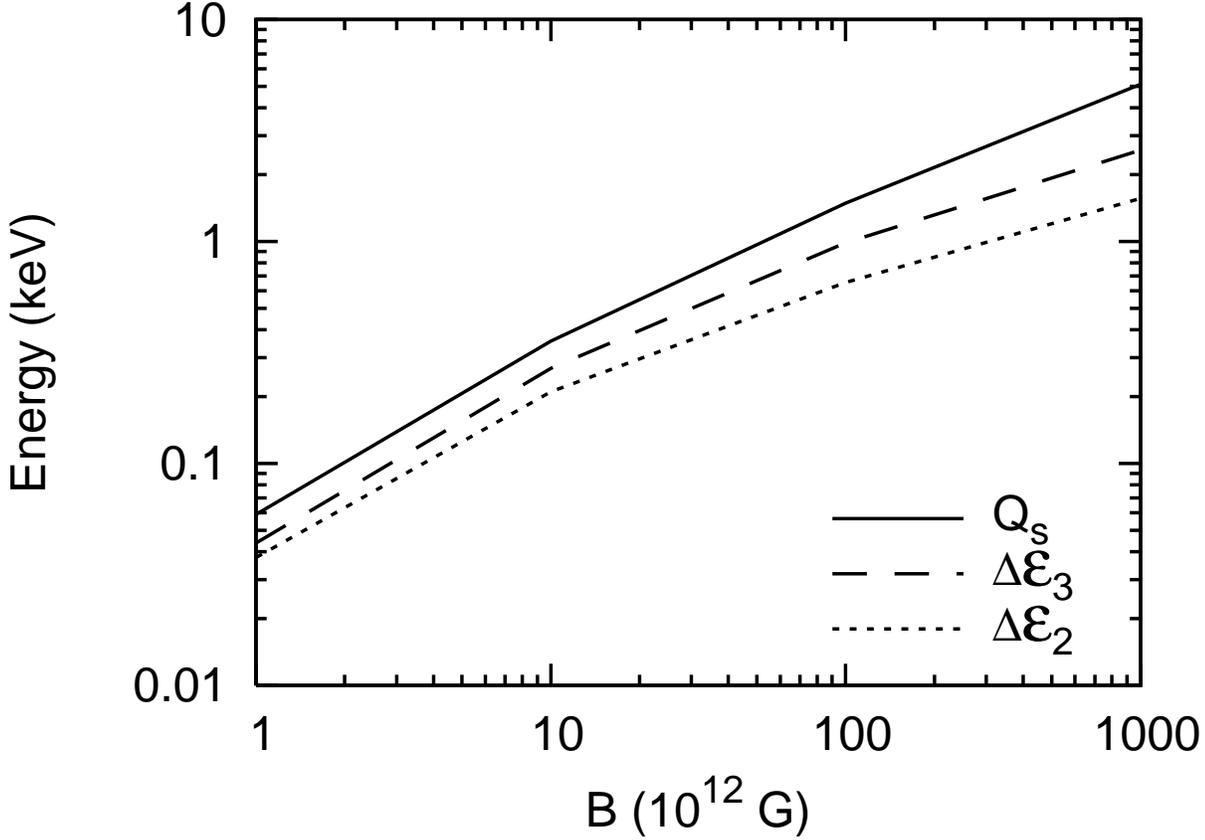}
\caption{Cohesive energy $Q_s={\cal E}_1-{\cal E}_s$ and molecular
energy difference $\Delta {\cal E}_N={\cal E}_N/N-{\cal E}_1$ for
helium as a function of the magnetic field strength.}
\label{HeEdfig}
\end{figure}

\begin{figure}
\includegraphics[width=6.5in]{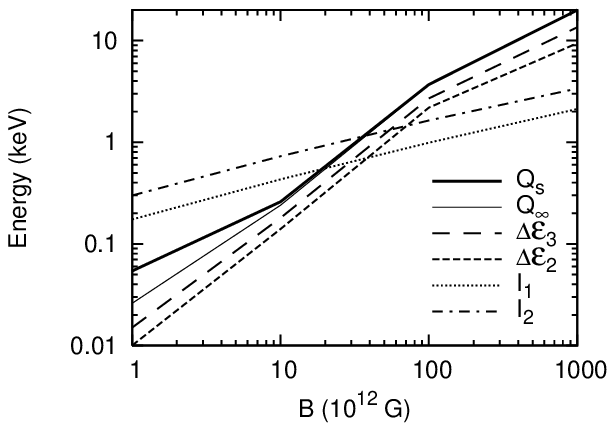}
\caption{Cohesive energy $Q_s={\cal E}_1-{\cal E}_s$ and molecular
energy difference $\Delta {\cal E}_N={\cal E}_N/N-{\cal E}_1$ for
carbon as a function of the magnetic field strength. The symbol
$Q_\infty$ represents the cohesive energy of a one-dimensional chain,
and $I_1$ and $I_2$ are the first and second ionization energies of
the C atom.}
\label{CEdfig}
\end{figure}

\begin{figure}
\includegraphics[width=6.5in]{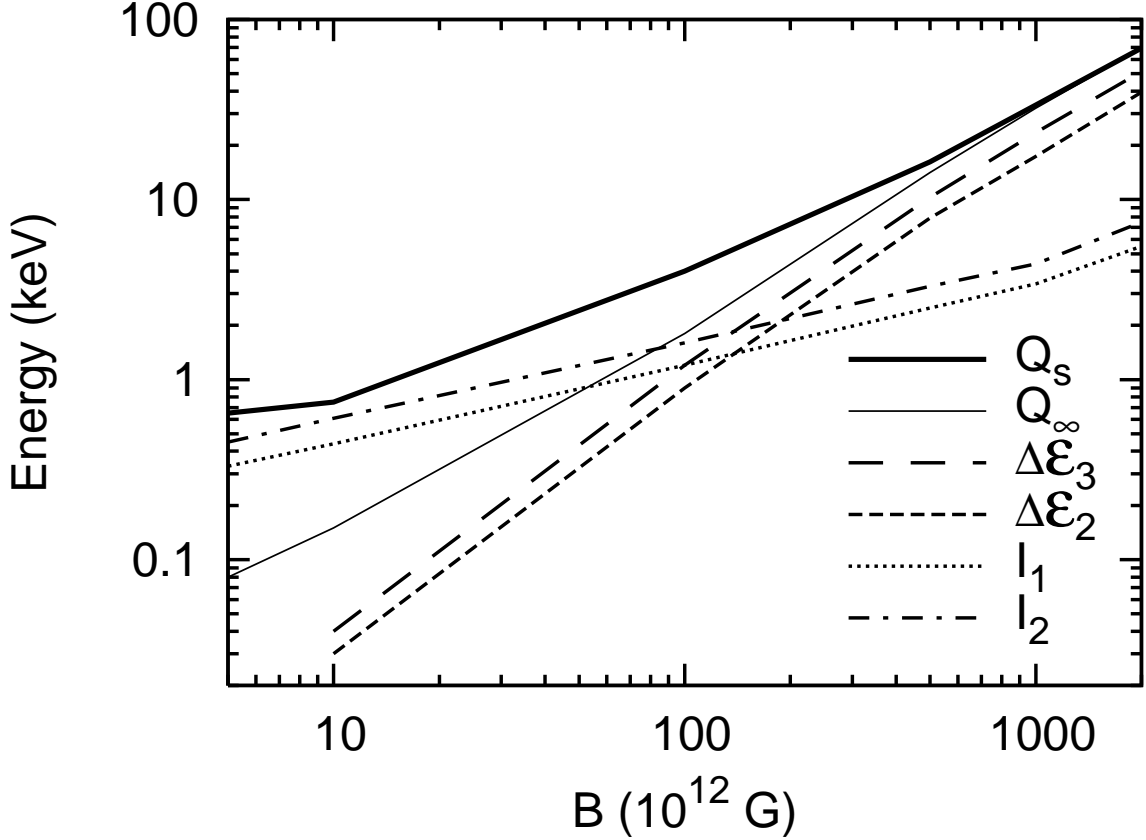}
\caption{Cohesive energy $Q_s={\cal E}_1-{\cal E}_s$ and molecular
energy difference $\Delta {\cal E}_N={\cal E}_N/N-{\cal E}_1$ for iron
as a function of the magnetic field strength. The symbol $Q_\infty$
represents the cohesive energy of a one-dimensional chain, and $I_1$
and $I_2$ are the first and second ionization energies of the Fe
atom. Below $5\times10^{12}$~G, our results for $Q_\infty$ and $Q_s$
become unreliable as $Q_\infty$ and $Q_s$ become very small and
approach numerical errors for ${\cal E}_N$ and ${\cal E}_s$.}
\label{FeEdfig}
\end{figure}

\begin{figure}
\includegraphics[width=6.5in]{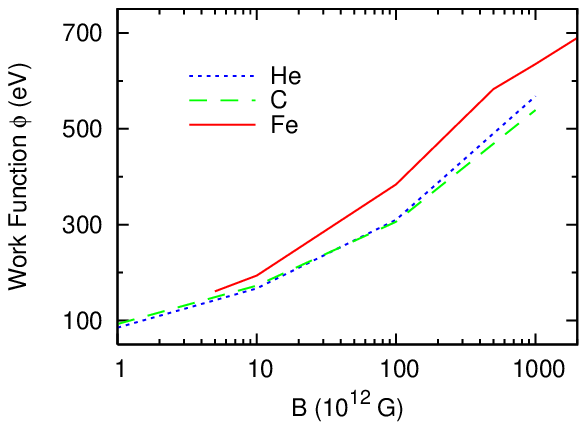}
\caption{Numerical result for the electron work function as a function
of the magnetic field strength, for He, C, and Fe infinite chains.}
\label{Wfig}
\end{figure}

Some numerical results from ML06a,b are provided in graphical form in
Figs.~\ref{HeEdfig}, \ref{CEdfig}, \ref{FeEdfig}, and \ref{Wfig} (see
ML06a,b for approximate scaling relations for different field ranges
based on numerical fits). Figure~\ref{HeEdfig} shows the cohesive
energies of condensed matter, $Q_s={\cal E}_1-{\cal E}_s$, and the
molecular energy differences, $\Delta {\cal E}_N={\cal E}_N/N-{\cal
E}_1$, for He, Fig.~\ref{CEdfig} for C, and Fig.~\ref{FeEdfig} for Fe;
here ${\cal E}_1$ is the atomic ground-state energy, ${\cal E}_N$ is
the ground-state energy of the He$_N$, C$_N$, or Fe$_N$ molecule, and
${\cal E}_s$ is the energy per cell of the zero-pressure 3D condensed
matter. Some relevant ionization energies for the atoms are also
shown. Figure~\ref{Wfig} shows the electron work functions $\phi$ for
condensed He, C, and Fe as a function of the field strength. We see
that the work function increases much more slowly with $B$ compared to
the simple free electron gas model [see Eq.~(\ref{fermieq})], and the
dependence on $Z$ is also weak. The results summarized here will be
used in Section~\ref{sect:cond} and Section~\ref{sect:polar} below.

\section{Condensation of Neutron Star Surfaces in Strong Magnetic Fields}
\label{sect:cond}

As seen from Figs.~\ref{HeEdfig}, \ref{CEdfig}, and \ref{FeEdfig}, the
cohesive energies of condensed matter increase with magnetic field. We
therefore expect that for sufficiently strong magnetic fields, there
exists a critical temperature $T_{\rm crit}$ below which a first-order
phase transition occurs between the condensate and the gaseous
vapor. This has been investigated in detail for hydrogen surfaces (see
\citealt{lai97,lai01}), but not for other surface compositions. Here
we consider the possibilies of such phase transitions
of He, C, and Fe surfaces.

A precise calculation of the critical temperature $T_{\rm crit}$ is
difficult. We can determine $T_{\rm crit}$ approximately by
considering the equilibrium between the condensed phase (labeled
``s'') and the gaseous phase (labeled ``g'') in the ultrahigh field
regime (where phase separation exists). The gaseous phase consists of
a mixture of free electrons and bound ions, atoms, and
molecules. Phase equilibrium requires the temperature, pressure and
the chemical potentials of different species to satisfy the conditions
(here we consider Fe as an example; He and C are similar)
\be
P_s = P_g = [2n({\rm Fe}^+) + 3n({\rm Fe}^{2+}) + \cdots + n({\rm Fe}) + n({\rm Fe}_2) + n({\rm Fe}_3) + \cdots]kT \,,
\ee
\be
\mu_s = \mu_e + \mu({\rm Fe}^+) = 2\mu_e + \mu({\rm Fe}^{2+}) = \cdots = \mu({\rm Fe}) = \frac{1}{2} \mu({\rm Fe}_2) = \frac{1}{3} \mu({\rm Fe}_3) = \cdots \,,
\ee
where we treat the gaseous phase as an ideal gas.
The chemical potential of the condensed phase is given by
\be
\mu_s = {\cal E}_s+P_s V_s \simeq {\cal E}_{s,0} \,,
\ee
where ${\cal E}_s$ is the energy per cell of the condensate and ${\cal E}_{s,0}$ is
the energy per cell at zero-pressure (we will label this simply as
${\cal E}_s$). We have assumed that the vapor pressure is sufficiently small
so that the deviation from the zero-pressure state of the condensate
is small; this is justified when the saturation vapor pressure $P_{\rm
sat}$ is much less than the critical pressure $P_{\rm crit}$ for phase
separation, or when the temperature is less than the critical
temperature by a factor of a few.

For nondegenerate electrons in a strong magnetic field the number
density is related to $\mu_e$ by
\ba
n_e & \simeq & \frac{1}{2\pi \rho_0^2} e^{\mu_e/kT} \sum_{n_L=0}^{\infty} g_{n_L} \exp\left(\frac{-n_L \hbar\omega_{ce}}{kT}\right) \int_{-\infty}^{\infty} \frac{dp_z}{h} \, \exp\left(\frac{-p_z^2}{2m_e kT}\right) \\
 & \simeq & \frac{1}{2\pi \rho_0^2 \lambda_{Te}} e^{\mu_e/kT} \tanh^{-1}\left(\frac{\hbar\omega_{ce}}{2kT}\right) \\
 & \simeq & \frac{1}{2\pi \rho_0^2 \lambda_{Te}} e^{\mu_e/kT} \,,
\ea
where $g_{n_L}=1$ for $n_L=0$ and $g_{n_L}=2$ for $n_L>0$ are the
Landau degeneracies, $\lambda_{Te}=(2\pi\hbar^2/m_e kT)^{1/2}$ is the
electron thermal wavelength, and the last equality applies for $kT \ll
\hbar\omega_{ce}$. The magnetic field length is $\rho_0 = (\hbar
c/eB)^{1/2}$. For atomic, ionic, or molecular Fe the number density is
given by
\ba
n({\rm Fe}_A) & \simeq & \frac{1}{h^3} e^{\mu_A/kT} \sum_i \exp\left(-\frac{{\cal E}_{A,i}}{kT}\right) \int d^3K \, \exp\left(\frac{-K^2}{2M_A kT}\right) \\
 & \simeq & \frac{1}{\lambda_{TA}^3} \exp\left(-\frac{{\cal E}_A-\mu_A}{kT}\right) Z_{\rm int}({\rm Fe}_A) \,,
\ea
with the internal partition function
\be
Z_{\rm int}({\rm Fe}_A) = \sum_i \exp\left(-\frac{\Delta {\cal E}_{A,i}}{kT}\right) \,.
\label{zinteq}
\ee
and $\Delta {\cal E}_{A,i}={\cal E}_{A,i}-{\cal E}_A$. Here, the
subscript $A$ represents the atomic, ionic, or molecular species whose
number density we are calculating (e.g., Fe$_2$ or Fe$^+$) and the sum
$\sum_i$ is over all excited states of that species. Also,
$\lambda_{Te}=(2\pi\hbar^2/M_A kT)^{1/2}$ is the Fe particle's thermal
wavelength, where $M_A = NAM$ is the total mass of the particle ($N$
is the number of ``atoms'' in the molecule, $A$ is the atomic mass
number, and $M=m_p+m_e$). The vector $\mathbf{K}$ represents the
center-of-mass momentum of the particle. Note that we have assumed
here that the Fe$_A$ particle moves across the field freely; this is a
good approximation for large $M_A$. The internal partition function
$Z_{\rm int}$ represents the effect of all excited states of the
species on the total density; in this work we will use the
approximation that this factor is the same for all species, and we
will estimate the magnitude of this factor later in this section.

The equilibrium condition $\mu_s=\mu({\rm Fe})$ for the process
${\rm Fe}_{s,\infty}+{\rm Fe} = {\rm Fe}_{s,\infty+1}$ yields the atomic
density in the saturated vapor:
\be
n({\rm Fe}) \simeq \left(\frac{AM kT}{2\pi \hbar^2}\right)^{3/2} \exp\left(-\frac{Q_s}{kT}\right) Z_{\rm int} \,,
\label{atomeq}
\ee
where $Q_s={\cal E}_1-{\cal E}_s$ is the cohesive energy of the
condensed Fe. The condition $N\mu_s=\mu({\rm Fe}_N)$ for the process
${\rm Fe}_{s,\infty}+{\rm Fe}_N = {\rm Fe}_{s,\infty+N}$ yields the
molecular density in the vapor:
\be
n({\rm Fe}_N) \simeq \left(\frac{NAM kT}{2\pi \hbar^2}\right)^{3/2} \exp\left(-\frac{S_N}{kT}\right) Z_{\rm int} \,,
\label{moleq}
\ee
where
\be
S_N = {\cal E}_N - N {\cal E}_s = N [Q_s - ({\cal E}_1-{\cal E}_N/N)]
\ee
is the ``surface energy'' and ${\cal E}_N/N$ is the energy per ion in the
molecule. The equilibrium condition $\mu({\rm Fe}^{n+})=\mu_e+\mu({\rm
Fe}^{(n+1)+})$ for the process $e+{\rm Fe}^{n+} = {\rm Fe}^{(n+1)+}$,
where Fe$^{n+}$ is the $n$th ionized state of Fe, yields the vapor
densities for the ions:
\be
n({\rm Fe}^+) n_e \simeq \frac{b}{2\pi a_0^2} \sqrt{\frac{m_e kT}{2\pi \hbar^2}} \exp\left(-\frac{I_1}{kT}\right) n({\rm Fe}) \,,
\label{ion1eq}
\ee
\be
n({\rm Fe}^{2+}) n_e \simeq \frac{b}{2\pi a_0^2} \sqrt{\frac{m_e kT}{2\pi \hbar^2}} \exp\left(-\frac{I_2}{kT}\right) n({\rm Fe}^+) \,,
\label{ion2eq}
\ee
and so on. Here, $b=B/B_0$ and $a_0$ is the Bohr radius, and
$I_n={\cal E}_{(n-1)+}-{\cal E}_{n+}$ represents the ionization energy
of the $n$th ionized state of Fe (i.e., the amount of energy required
to remove the $n$th electron from the atom when the first $n-1$
electrons have already been removed). The total electron density in
the saturated vapor is
\be
n_e = n({\rm Fe}^+)+2n({\rm Fe}^{2+})+\cdots \,.
\label{eleceq}
\ee
The number densities of electrons [Eq.~(\ref{eleceq})] and ions [e.g.,
Eqs.~(\ref{ion1eq}) and (\ref{ion2eq})] must be found
self-consistently, for all ion species that contribute significantly
to the total vapor density. The total mass density in the vapor is
calculated from the number densities of all of the species discussed
above, using the formula
\be
\rho_g = AM\left[n({\rm Fe})+2n({\rm Fe}_2)+\cdots+n({\rm Fe}^+)+n({\rm Fe}^{2+})+\cdots\right] \,.
\ee

\begin{figure}
\begin{center}
\begin{tabular}{cc}
\resizebox{3in}{!}{\includegraphics{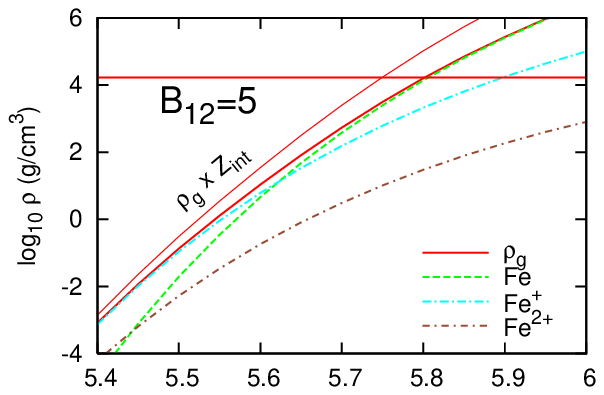}} & \\
\resizebox{3in}{!}{\includegraphics{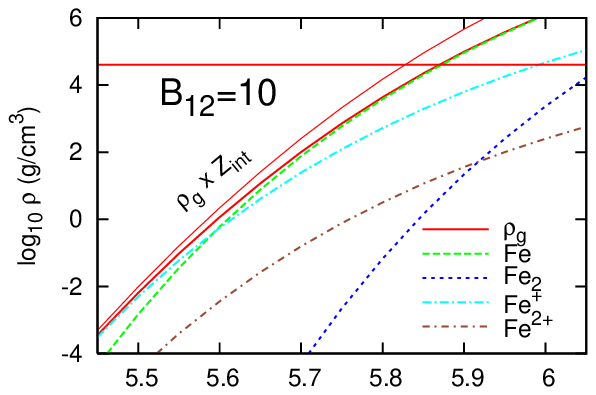}} &
\resizebox{3in}{!}{\includegraphics{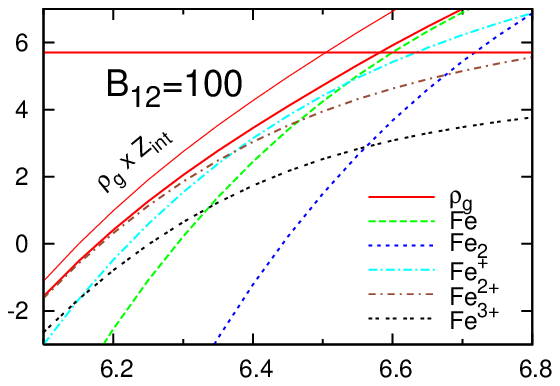}} \\
\resizebox{3in}{!}{\includegraphics{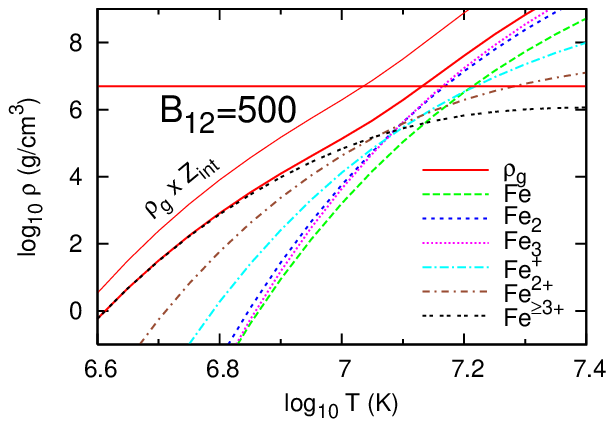}} &
\resizebox{3in}{!}{\includegraphics{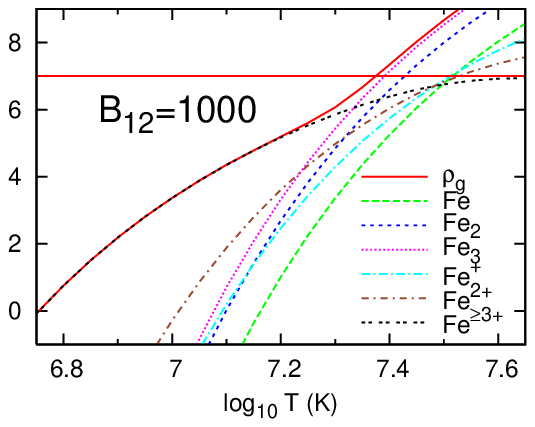}} \\
\end{tabular}
\caption{The mass densities of various atomic/ionic/molecular species
and the total density ($\rho_g$) of the vapor in phase equilibrium
with the condensed iron surface. The five panels are for different
field strengths, $B_{12}=5,10,100,500,1000$. The horizontal lines give
the densities of the condensed phase, $\rho_s$. All the vapor density
curves are calculated assuming $Z_{\rm int}=1$, except for the curve
marked by ``$\rho_g\times Z_{\rm int}$'', for which the total vapor
density is calculated taking into account the nontrivial internal
partition functions of various species. The critical temperature
$T_{\rm crit}$ for phase separation is set by the condition
$\rho_g=\rho_s$.}
\label{FeGasfig}
\end{center}
\end{figure}

\begin{figure}
\begin{center}
\begin{tabular}{cc}
\resizebox{3in}{!}{\includegraphics{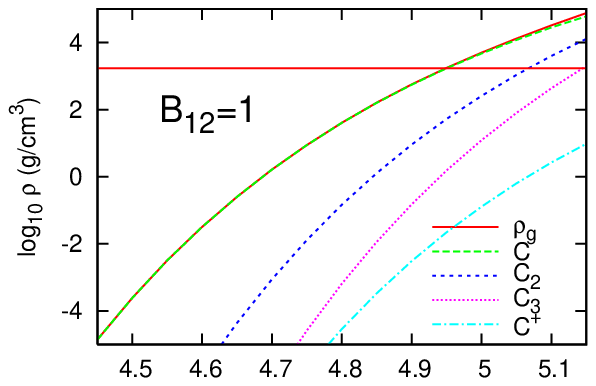}} &
\resizebox{3in}{!}{\includegraphics{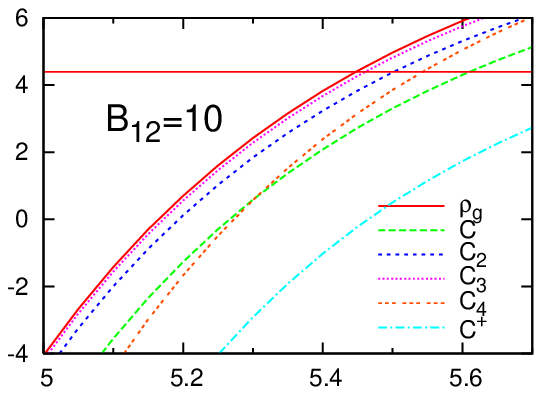}} \\
\resizebox{3in}{!}{\includegraphics{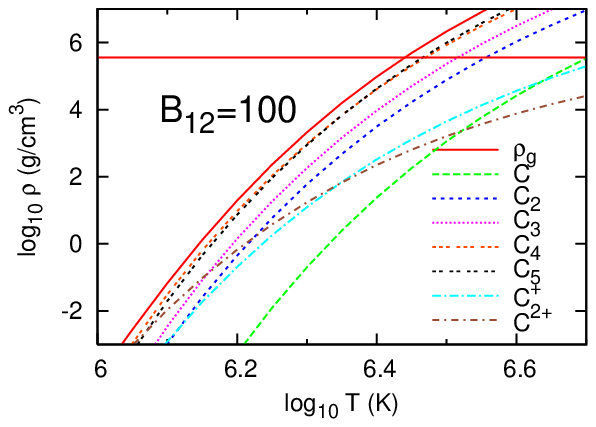}} &
\resizebox{3in}{!}{\includegraphics{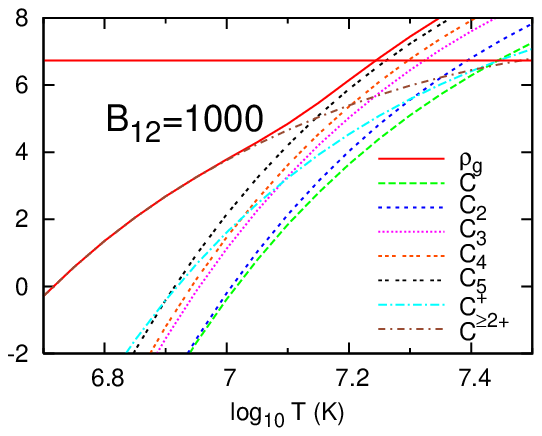}} \\
\end{tabular}
\caption{The mass densities of various atomic/ionic/molecular species
and the total density ($\rho_g$) of the vapor in phase equilibrium
with the condensed carbon surface. The four panels are for different
field strengths, $B_{12}=1,10,100,1000$. The horizontal lines give the
densities of the condensed phase, $\rho_s$. All the vapor density
curves are calculated assuming $Z_{\rm int}=1$. The critical
temperature $T_{\rm crit}$ for phase separation is set by the
condition $\rho_g=\rho_s$.}
\label{CGasfig}
\end{center}
\end{figure}

Figure~\ref{FeGasfig} (for Fe) and Fig.~\ref{CGasfig} (for C) show the
the densities of different atomic/molecular species in the saturated
vapor in phase equilibirum with the condensed matter for different
temperatures and field strengths. These are computed using the values
of ${\cal E}_N/N$, ${\cal E}_s$, and ${\cal E}_{n+}$ presented in
ML06a,b and depicted in Figs.~\ref{CEdfig} and \ref{FeEdfig}. As
expected, for sufficiently low temperatures, the total gas density in
the vapor is much smaller than the condensation density, and thus
phase separation is achieved. The critical temperature $T_{\rm crit}$,
below which phase separation between the condensate and the gaseous
vapor occurs, is determined by the condition $\rho_s = \rho_g$. We
find that for Fe:
\be
T_{\rm crit} \simeq 6\times10^5,~7\times10^5,~3\times10^6,~10^7,
~2\times10^7~{\rm K}\quad
{\rm for}~~B_{12}=5,~10,~100,~500,~1000,
\ee
for C:
\be
T_{\rm crit} \simeq 9\times10^4,~3\times10^5,~3\times10^6,~2\times10^7~{\rm K}\quad
{\rm for}~~B_{12}=1,~10,~100,~1000.
\ee
and for He:
\be
T_{\rm crit} \simeq 8\times10^4,~3\times10^5,~2\times10^6,~9\times10^6~{\rm K}\quad
{\rm for}~~B_{12}=1,~10,~100,~1000.
\ee
In terms of the cohesive energy, these results can be approximated by
\be
kT_{\rm crit} \sim 0.08\,Q_s \,.
\ee

Note that in our calculations for the iron vapor density at
$B_{12}=5$-$500$ we have estimated the magnitude of the internal
partition function factor $Z_{\rm int}$; the modified total density
curves are marked on these figures as ``$\rho_g \times Z_{\rm int}$''.
To estimate $Z_{\rm int}$ we use Eq.~(\ref{zinteq}) with a cutoff to
the summation above some energy. For $B_{12}=5,10,100$, and $500$ we
calculate or interpolate the energies for all excited states of atomic
Fe with energy below this cutoff, in order to find $Z_{\rm int}$.
The energy cutoff is necessary because the highly excited states
become unbound (ionized) due to finite pressure and should not be
included in $Z_{\rm int}$ (otherwise $Z_{\rm int}$ would diverge). In
principle, the cutoff is determined by requiring the effective size of
the excited state to be smaller than the inter-particle space in the
gas, which in turn depends on density. In practice, we choose the
cutoff such that the highest excited state has a binding energy
$|{\cal E}_{A,i}|$ significantly smaller than the ground-state binding energy
$|{\cal E}_A|$ (typically 30\% of it). As an approximation, we also assume
that the internal partitions for Fe$_N$ molecules and ions have the
same $Z_{\rm int}$ as the Fe atom. Despite the crudeness of our
calculation of $Z_{\rm int}$, we see from Fig.~\ref{FeGasfig} that the
resulting $T_{\rm crit}$ is only reduced by a few tens of a percent
from the $T_{\rm crit}$ value assuming $Z_{\rm int}=1$.

We note that our calculation of the saturated vapor density is very
uncertain around $T \sim T_{\rm crit}$, since
Eqs.~(\ref{atomeq})\,--\,(\ref{ion2eq}) are derived for $\rho_g \ll
\rho_s$ while the critical temperature of the saturated vapor density
is found by setting $\rho_s = \rho_g$. However, since the vapor
density decreases rapidly as $T$ decreases, when the temperature is
below $T_{\rm crit}/2$ (for example), the vapor density becomes much
less than the condensation density and phase transition is
unavoidable. When the temperature drops below a fraction of $T_{\rm
crit}$, the vapor density becomes so low that the optical depth of the
vapor is negligible and the outermost layer of the neutron star then
consists of condensed matter. The radiative properties of such
condensed phase surfaces have been studied using a simplified
treatment of the condensed matter (see \citealt{vanadelsberg05} and
references therein).

\section{Polar Vacuum Gap Acclerators in Pulsars and Magnetars}
\label{sect:polar}

A rotating, magnetized neutron star is surrounded by a magnetosphere
filled with plasma. The plasma is assumed to be an excellent
conductor, such that the charged particles move to screen out any
electric field parallel to the local magnetic field. The
corresponding charge density is given by (\citealt{goldreich69})
\be
\rho_{GJ} \simeq -\frac{\mathbf{\Omega} \cdot \mathbf{B}}{2\pi c} 
\label{rhoGJeq}
\ee
where $\mathbf{\Omega}$ is the rotation rate of the neutron star.

The Goldreich-Julian density assumes that charged particles are always
available. This may not be satisfied everywhere in the
magnetosphere. In particular, charged particles traveling outward
along the open field lines originating from the polar cap region of
the neutron star will escape beyond the light cylinder. To maintain
the required magnetosphere charge density these particles have to be
replenished by the stellar surface. If the surface temperature and
cohesive strength are such that the required particles are tightly
bound to the stellar surface, those regions of the polar cap through
which the charged particles are escaping will not be replenished. A
vacuum gap will then develop just above the polar cap (e.g.,
\citealt*{ruderman75,cheng80,usov96,zhang00,gil03}). In this vacuum gap
zone the parallel electric field is no longer screened and particles
are accelerated across the gap until vacuum breakdown (via pair
cascade) shorts out the gap. Such an acceleration region can have an
important effect on neutron star emission processes. We note that in
the absence of a vacuum gap, a polar gap acceleration zone based on
space-charge-limited flow may still develop (e.g.,
\citealt{arons79,harding98,muslimov03}).

In this section we determine the conditions required for the vacuum
gap to exist using our results summarized in
Section~\ref{sect:cohes}. The cohesive energy and electron work
function of the condensed neutron star surface are obviously the key
factors. We examine the physics of particle emission from condensed
surface in more detail than considered previously.

\subsection{Particle Emission From Condensed Neutron Star Surfaces}
\label{sect:density}

We assume that the NS surface is in the condensed state, i.e., the
surface temperature $T$ is less than the critical temperature $T_{\rm
crit}$ for phase separation (see Section~\ref{sect:cond}). (If
$T>T_{\rm crit}$, the surface will be in gaseous phase and a vacuum
gap will not form.) We shall see that in order for the surface not to
emit too large a flux of charges to the magnetosphere (a necessary
condition for the vacuum gap to exist), an even lower surface
temperature will be required.

\subsubsection{Electron Emission}
\label{sect:elec}

For neutron stars with $\mathbf{\Omega}\cdot\mathbf{B}_p > 0$, where
$\mathbf{B}_p$ is the magnetic field at the polar cap, the
Goldreich-Julian charge density is negative at the polar cap, thus
surface electron emission (often called thermionic emission in solid
state physics; \citealt{ashcroft76}) is relevant. Let ${\cal F}_e$ be
the number flux of electrons emitted from the neutron star
surface. The emitted electrons are accelerated to relativistic speed
quickly, and thus the steady-state charge density is $\rho_e=-e{\cal
F}_e/c$. For the vacuum gap to exist, we require $|\rho_e|<|\rho_{\rm
GJ}|$. (If $|e{\cal F}_e/c|>|\rho_{\rm GJ}|$, the charges will be
rearranged so that the charge density equals $\rho_{\rm GJ}$.)

To calculate the electron emission flux from the condensed surface,
we assume that these electrons behave like a free
electron gas in a metal, where the energy barrier they must overcome
is the work function of the metal. In a strong magnetic field, the electron 
flux is given by 
\be
{\cal F}_e = \int_{p_{\rm min}}^\infty f(\epsilon) \frac{p_z}{m_e} \frac{1}{2\pi \rho_0^2}\frac{dp_z}{h} \,,
\ee
where $p_{\rm min}=\sqrt{2m_e|U_0|}$, $U_0$ is the potential energy of
the electrons in the metal, $\epsilon=p_z^2/(2m_e)$ is the electron
kinetic energy, and
\be
f(\epsilon) = \frac{1}{e^{(\epsilon-\mu'_e)/kT}+1}
\ee
is the Fermi-Dirac distribution function with $\mu'_e$ the electron
chemical potential (excluding potential energy). Integrating this
expression gives
\be
{\cal F}_e = \frac{kT}{2\pi h \rho_0^2} \ln\left[1+e^{-\phi/kT}\right] \simeq \frac{kT}{2\pi h \rho_0^2} e^{-\phi/kT}\,,
\ee
where $\phi \equiv |U_0|-\mu'_e$ is the work function of the condensed
matter and the second equality assumes $\phi \gg kT$.
The steady-state charge density supplied by the surface is then
\be
\rho_e = -\frac{e}{c}{\cal F}_e = \rho_{GJ} \exp{(C_e-\phi/kT)} \,,
\label{rhoeeq}
\ee
with
\be
C_e = \ln\left(\frac{e}{c}\frac{kT}{2\pi h \rho_0^2|\rho_{GJ}|}\right) \simeq 31+\ln{(P_0 T_6)} \sim 30 \,,
\label{Ceeq}
\ee
where $T_6=T/(10^6~{\rm K})$ and $P_0$ is the spin period in units of
1~s. For a typical set of pulsar parameters (e.g., $P_0=1$ and
$T_6=0.5$) $C_e \sim 30$, but $C_e$ can range from 23 for millisecond
pulsars to 35 for some magnetars. Note that the requirement $\phi \gg
kT$ is automatically satified here when $|\rho_e|$ is less than
$|\rho_{GJ}|$. The electron work function was calculated in ML06b and
is depicted in Fig.~\ref{Wfig}.

\subsubsection{Ion Emission}
\label{sect:ion}

For neutron stars with $\mathbf{\Omega}\cdot\mathbf{B}_p < 0$, the
Goldreich-Juliam charge above the polar cap is positive, so we are
interested in ion emission from the surface.  Unlike the electrons,
which form a relatively free-moving gas within the condensed matter,
the ions are bound to their lattice sites.\footnote{The freezing
condition is easily satisfied for condensed matter of heavy elements
(see \citealt{vanadelsberg05}).} To escape from the surface, the ions
must satisfy three conditions. First, they must be located on the
surface of the lattice. Ions below the surface will encounter too much
resistance in trying to move through another ion's cell. Second, they
must have enough energy to escape as unbound ions. This binding energy
that must be overcome will be labeled ${\cal E}_B$. Third, they must be
thermally activated. The energy in the lattice is mostly transferred
by conduction, so the ions must wait until they are bumped by atoms
below to gain enough energy to escape.

Consider the emission of ions with charge $Z_n e$ from the neutron
star surface (e.g., Fe$^+$ would have $Z_n=1$). The rate of collisions
between any two ions in the lattice is approximately equal to the
lattice vibration frequency $\nu_i$, which can be estimated from
\be
\nu_i=\frac{1}{2\pi}\left(\Omega_p^2+\omega_{ci}^2\right)^{1/2} \,,
\ee
where $\Omega_p=\left(4\pi Z^2e^2n_i/m_i\right)^{1/2}$ is the ion
plasma (angular) frequency and $\omega_{ci}=ZeB/(m_ic)$ is the ion
cyclotron frequency ($m_i=Am_p$).
Not all collisions will lead to ejection of ions from
the surface, since an energy barrier ${\cal E}_B$ must be overcome. Thus each
surface ion has an effective emission rate of order
\be
\chi = \nu_i e^{-{\cal E}_B/kT} \,.
\ee
The energy barrier ${\cal E}_B$ for ejecting ions of charge $Z_n e$ is
equivalent to the energy required to release a neutral atom from the
surface and ionize it, minus the energy gained by returning the
electron to the surface (e.g., \citealt{tsong90}). Thus
\be
{\cal E}_B = Q_s + \sum_{i=1}^{Z_n}I_i - Z_n\phi \,,
\label{EBeq}
\ee
where $Q_s>0$ is the cohesive energy, $I_i>0$ is the $i$th ionization
energy of the atom (so that $\sum_{i=1}^{Z_n}I_i$ is the energy
required to remove $Z_n$ electrons from the atom), and $\phi>0$ is the
electron work function. The surface density of ions is $n_i r_i$,
where $r_i$ is the mean spacing between ions in the solid. Thus the
emission flux of $Z_n$-ions is
\be
{\cal F}_i = \nu_i n_i r_i e^{-{\cal E}_B/kT} \,.
\ee
The steady-state $Z_n$-ion number density supplied by the surface is then
\be
\rho_i = \frac{Z_n e}{c}{\cal F}_i = \rho_{GJ} \exp(C_i-{\cal E}_B/kT)\,,
\label{rhoieq}
\ee
with
\ba
C_i & = & \ln\left(\frac{Z_n e \nu_i n_i r_i}{c \rho_{GJ}}\right) \nonumber\\
 & \simeq & 34+\ln\left\{Z_n Z A^{-1/2} n_{28}^{3/2} (r_i/a_0) B_{12}^{-1} P_0 \sqrt{1+5.2\times10^{-3} A^{-1} B_{12}^2 n_{28}^{-1}} \right\} \sim \mbox{27--33} \,,
\label{Cieq}
\ea
where $n_{28}=n_i/(10^{28}~{\rm cm}^{-3})$. For a typical set of
pulsar parameters (e.g., $B_{12}=1$ and $P_0=1$) $C_i \sim 27$, but
$C_i$ can be as large as 33 for magnetars with $B_{12}=1000$ and $P_0=8$.

All the quantities in ${\cal E}_B$ were calculated in ML06b (see
Figs.~\ref{CEdfig} and \ref{FeEdfig}). We find that the emission of
singly-ionized atoms ($Z_n=1$) is most efficient, as ${\cal E}_B$ is
signficantly lower for $Z_n=1$ than for $Z_n>1$ ($\sum_{i=1}^{Z_n}I_i$
grows much faster with $Z_n$ than $Z_n \phi$ does).

\subsubsection{Effect of Electric Field on Charge Emission}
\label{sect:electric}

The discussion in Sections~\ref{sect:elec} and \ref{sect:ion} includes
only thermal emission of charged particles from the condensed
surface. A strong electric field, of order $E_s\sim \Omega B R/c$, may
be present. Since this electric field is much less than the
characteristic field $\sim e/r_i^2$ inside the condensed matter (where
$r_i$ is the mean particle separation), this field cannot directly rip
charges off the surface. Nevertheless, the electric field may enhance
the thermal emission of charge particles. We now estimate the
magnitude of this effect.

In the presence of a vacuum gap, the electric field $E_s$ at
the stellar surface points outward ($E_s>0$) for stars with
$\mathbf{\Omega}\cdot\mathbf{B}_p < 0$ and inward ($E_s<0$) for
stars with $\mathbf{\Omega}\cdot\mathbf{B}_p >0$.  A charge $Q$ moved
to some small height $z$ above the surface gains a potential energy
given by $U=-Q^2/(4z)-QE_s z$, where the first term is due to
the interaction between the charge and the perfectly conducting metal
surface, and the second term is due to the external field.\footnote{In
the vacuum gap, the electric field is not exactly uniform, but since
the maximum $U$ is attained at a rather small height compared to the
gap thickness, this nonuniformity is unimportant for our consideration
here.} The potential reaches a maximum value
\be
U_{\rm max}= - |Q|^{3/2}|E_s|^{1/2}
\ee
at the height $z=|Q/4E_s|^{1/2}$.
Thus, compared to the $E_s=0$ case, the energy barrier 
for particle emission is now reduced by the amount $U_{\rm max}$.

Combining this consideration with the results of
Sections~\ref{sect:elec} and \ref{sect:ion}, we find that steady-state
charge density due to electron surface emission (for $\mathbf{\Omega}
\cdot \mathbf{B_p}>0$ stars) is
(cf. \citealt{jessner01})
\be
\rho_e=\rho_{GJ}\exp [C_e-(\phi-e^{3/2}|E_s|^{1/2})/kT],
\label{rhoieq2}
\ee
and the steady-state charge density due to ion surface emission 
(for $\mathbf{\Omega} \cdot \mathbf{B_p}<0$ stars) is 
\be
\rho_i=\rho_{GJ}\exp [C_i-({\cal E}_B-(Z_ne)^{3/2}|E_s|^{1/2})/kT].
\label{rhoeeq2}
\ee
For $E_s\sim \Omega B R/c$, we have 
$e^{3/2}|E_s|^{1/2}\sim 10$~eV. This is typically much smaller 
than either $\phi$ or ${\cal E}_B$.

\subsection{Conditions for Gap Formation}
\label{sect:final}

No vacuum gap will form if the electrons or ions are able to fill the
magnetosphere region above the polar cap with the required
Goldreich-Julian density; i.e., the vacuum gap will cease to exist
when $\rho_e=\rho_{GJ}$ or $\rho_i=\rho_{GJ}$. From
Eqs.~(\ref{rhoeeq2}) and (\ref{rhoieq2}) we can see that no polar gap
will form if
\be
\phi-e^{3/2}|E_s|^{1/2} < C_e kT \sim 3 T_6 \mbox{ keV}
\label{finaleeq}
\ee
for a negative polar magnetosphere ($\mathbf{\Omega}\cdot\mathbf{B}_p
> 0$), and
\be
{\cal E}_B-(Z_n e)^{3/2}|E_s|^{1/2} < C_i kT \sim 3 T_6 \mbox{ keV}
\label{finalieq}
\ee
for a positive polar magnetosphere ($\mathbf{\Omega}\cdot\mathbf{B}_p
< 0$). [For the exact expressions for $C_e$ and $C_i$ see
Eqs.~(\ref{Ceeq}) and (\ref{Cieq}).]

\begin{figure}
\includegraphics[width=6.5in]{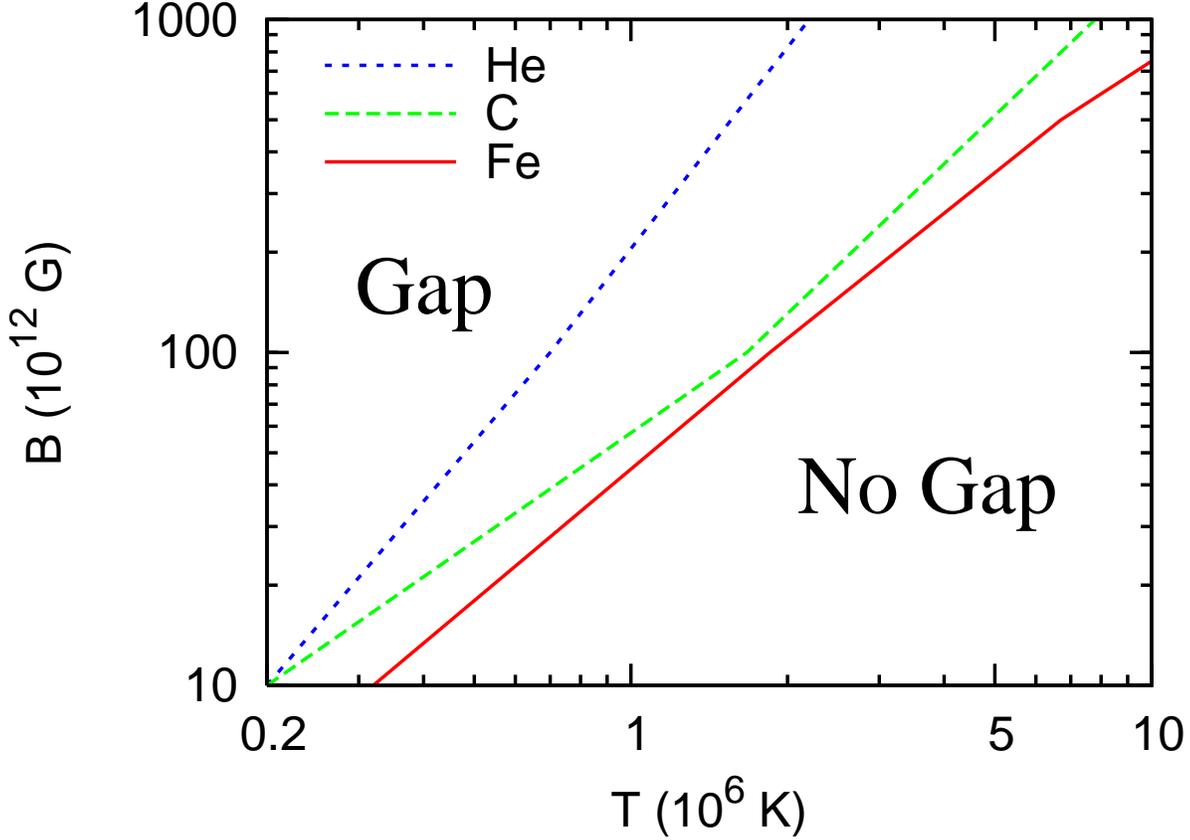}
\caption{The condition for the formation of a vacuum gap above
condensed helium, carbon, and iron neutron star surfaces, when the
magnetosphere is positive over the poles
($\mathbf{\Omega}\cdot\mathbf{B}_p<0$).}
\label{gapfig}
\end{figure}

For neutron stars in general, the electron work function $\phi$ is
much less than $C_ekT \sim 3 T_6$~keV (see Fig.~\ref{Wfig}), so
electrons can easily escape from the condensed surface. No gap forms
for a negative polar magnetosphere under neutron star surface
conditions. (This is contrary to the conclusions of
\citealt{usov96} and \citealt{gil03}.) The ion binding energy ${\cal
E}_B$ [given by Eq.~(\ref{EBeq})], on the other hand, can be larger
than $C_ikT \sim 3 T_6$~keV under certain neutron star surface
conditions (see Figs.~\ref{HeEdfig}, \ref{CEdfig}, and
\ref{FeEdfig}). Ions can tightly bind to the condensed surface and a
polar gap can form under these conditions. Figure~\ref{gapfig} shows
the critical temperature (determined by ${\cal E}_B = C_i kT$) below
which a vacuum gap can form for the Fe, C, and He surfaces.

\section{Vacuum Gap Acclerators: Pair Cascades and the Pulsar Death Line/Boundary}
\label{sect:death}

Pair cascading in the magnetosphere of a pulsar is an essential
ingredient for its radio emission (e.g., \citealt{melrose04}).  The
pair cascade involves: (a) acceleration of primary particles by an
electric field parallel to the magnetic field; (b) gamma ray emission
by the accelerated particles moving along the magnetic field lines
(either by curvature radiation or inverse Compton upscattering of
surface photons); (c) photon decay into pairs as the angle between the
photon and the field line becomes sufficiently large.
To initiate the cascade an acceleration region is required;
the characteristics of this particle accelerator determine whether pulsar
emission can operate or not (the so-called ``pulsar death line'';
e.g., \citealt{ruderman75,arons00,zhang00,hibschman01}). Depending on the
boundary condition at the neutron star surface, there are two types of
polar gap accelerators: If charged particles are strongly bound to the
neutron star surface by cohesive forces, a vacuum gap develops
directly above the surface, with height $h$ much less than the stellar
radius \citep{ruderman75}; if charged particles can be
freely extracted from the surface, a more extended
space-charge-limited-flow (SCLF) type accelerator develops due to
field line curvatures \citep{arons79} and the relativistic frame
dragging effect (e.g., \citealt{muslimov92}). Because the cohesive
strength of matter at $B\sim 10^{12}$~G was thought to be negligible
(based on the result of \citealt{neuhauser87}), most theoretical works
in recent years have focused on the SCLF models (e.g.,
\citealt{arons00,muslimov03,muslimov04}).

Our results in Section~\ref{sect:polar} show that for sufficiently
strong magnetic fields and/or low surface temperatures, a vacuum gap
accelerator can form. Such a vacuum gap may be particularly relevant
for the so-called high-B radio pulsars, which have inferred magnetic
fields similar to those of magnetars
(e.g., \citealt{mclaugh05,burgay06}). In this section we discuss the
conditions under which a vacuum gap will be an effective generator of
pulsar emission. As discussed in Section~\ref{sect:polar}, since
electrons are weakly bound to the condensed stellar surface, such a
vacuum gap is possible only for pulsars with
$\mathbf{\Omega}\cdot\mathbf{B}_p < 0$ (as suggested in the original
Ruderman-Sutherland model).

Our analysis is similar to the original Ruderman-Sutherland model,
except that we extend our discussion of the cascade physics to the
magnetar field regime, which introduces some corrections to previous
works (e.g., \citealt{ruderman75,usov96}). We also consider
photon emission due to inverse Compton scattering, in addition to
curvature radiation, in the cascade
(cf.~\citealt{zhang97,zhang00,hibschman01}).

\subsection{Acceleration Potential}

When the temperature drops below the critical value given in
Section~\ref{sect:polar}, the charge density above the polar cap
decreases quickly below $\rho_{\rm GJ}$, and a vacuum gap results. In
the vacuum region just above the surface ($0\le z\ll R$), the parallel
electric field satisfies the equation $dE_\parallel/dz\simeq
-4\pi \rho_{GJ}$. The height of the gap $h~(\ll R)$ is determined by
vacuum breakdown due to pair cascade, which shorts out the electric
field above the gap (i.e., $E_\parallel=0$ for $z\ge h$). Thus
the electric field in the gap is
\be
E_\parallel \simeq \frac{2 \Omega B_p}{c}(h-z),
\label{efieldeq}
\ee
where $B_p=b_dB_p^d$ is the actual magnetic field at the pole, and differs
from the dipole field $B_p^d$ by a factor $b_d\ge 1$.
The potential drop across the gap is then
\be
\Delta \Phi = \frac{\Omega B_p}{c}h^2=b_d\frac{\Omega B_p^d}{c}h^2.
\label{eq:Phidrop}
\ee
With this potential drop, the electrons and positrons can be
accelerated to a gamma factor 
\be
\gamma_m = \frac{e \Delta \Phi}{m_e c^2}  
=  5.43\times10^6 \beta_Q h_3^2 P_0^{-1}
=1.23\times10^5 b_d B_{12} h_3^2 P_0^{-1} \,,
\label{gammaeq}
\ee
where $\beta_Q=B_p/B_Q$ (with $B_Q=m_e^2 c^3/e\hbar = 
4.414\times10^{13}$~G the QED field), $B_{12}=B_p^d/(10^{12}~{\rm G})$,
$h_3=h/(10^3~{\rm cm})$ and $P_0$ is the spin period in units of 1~s.
The voltage drop across the gap can be no larger than the voltage drop
across the polar cap region 
$\Delta \Phi_{max}\simeq (\Omega B_p/2c)(r_{p_+})^2 =
(\Omega B_p^d/2c)(r_{p_+}^d)^2$, where $r_{p_+}=
r_{p+}^d/b_d^{1/2}$ is the radius of the 
polar cap through which a net postive current flows:
\be
r_{p_+}^d =\left(\frac{2}{3}\right)^{3/4} R\left(\frac{\Omega
R}{c}\right)^{1/2}.
\ee
Thus the gap height is limited from above by 
\be
h_{\rm max} \simeq \frac{r_{p_+}^d}{\sqrt{2b_d}} = 7.54\times10^3
\,b_d^{-1/2} P_0^{-1/2} \mbox{ cm,}
\label{hmaxeq}
\ee
where we have adopted $R=10$~km.

The above equations are for an aligned rotator. For an oblique rotator
(where the magnetic dipole axis is inclined relative to the rotation
axis), the voltage drop across the polar cap region is larger, of
order $(\Omega B_p/2c)Rr_{p_+}$. But as discussed in Appendix A, the
acceleration potential across the vacuum gap is still limited from
above by $\Delta \Phi_{max}\sim (\Omega B_p/2c)r_{p_+}^2$.

\subsection{Requirements for Gap Breakdown}

There are two requirements for the breakdown of a vacuum gap. 
First, the photons must be able to create electron-positron pairs 
within the gap, i.e., the mean free path of photon pair-production 
is less than the gap height:
\be
l_{\rm ph} < h \,.
\label{firstceq}
\ee
Second, the electrons and positrons must be accelerated over the
gap potential and produce at least several photons
within the gap. If on average only one
photon is emitted with the required energy for each electron-positron
pair, for instance, then the number of charged particles produced in
the gap will grow very slowly and the gap will not break down
completely. Therefore, we must have
\be
N_{\rm ph} > \lambda \,,
\label{secondceq}
\ee
where $N_{\rm ph}$ is the number of photons emitted within the gap by each
electron or positron, and $\lambda$ is a number of order $1$--$10$.

\subsection{Pair Production}

The threshold of pair production for a photon with energy $\epsilon$ is 
\be
\frac{\epsilon}{2m_ec^2}\sin\theta > 1\,,
\label{threseq}
\ee
where $\theta$ is the angle of intersection of the photon and the
magnetic field. Suppose a photon is emitted at an angle $\theta_e$.
After the photon travels a distance $z$, the intersection angle will
grow as $z/{\cal R}_c$, where ${\cal R}_c$ is the local radius of
curvature of the polar magnetic field line. Thus the typical
intersection angle (for a photon crossing the entire gap) is
\be
\sin\theta \simeq \theta\simeq \frac{h}{{\cal R}_c}+\theta_e.
\label{sineq}
\ee
For a pure dipole field, the curvature radius is of order 
$(Rc/\Omega)^{1/2} \simeq 10^8 P_0^{1/2}$~cm, but a more complex
field topology at the polar cap could reduce ${\cal R}_c$ to as small as 
the stellar radius.

In the weak-field regime, 
when the threshold condition is well-satisfied (so that the pairs are
produced in highly excited Landau levels), the mean free path 
is given by \citep{erber66}
\be
l_{\rm ph} \simeq \frac{4.4 a_0}{\beta_Q \sin\theta} \exp\left(\frac{4}{3\chi}
\right), \quad {\rm with}~~
\chi = \frac{\epsilon}{2m_ec^2}\beta_Q\, \sin\theta \,,
\label{lpheq}
\ee
where $a_0=\hbar^2/(m_ec^2)$ is the Bohr radius. The condition $l_{\rm ph}<h$
implies $\chi\ga 1/15$ for typical parameters \citep{ruderman75}.
For stronger magnetic fields ($\beta_Q\ga 0.1-0.2$), the pairs tend to be
produced at lower Landau levels. Using the general expression for the
pair production rate (e.g., \citealt{daugherty83}), one can 
check that if the threshold condition Eq.~(\ref{threseq}) is satisfied, 
the pair-production optical depth across the gap would also be greater 
than unity [for $\beta_Q=0.1$, the optical depth $\tau$ is unity when
$\epsilon/(2m_ec^2)\sin\theta > 1.05$, and by $\beta_Q=0.2$, $\tau=1$
when $\epsilon/(2m_ec^2)\sin\theta > 1+10^{-7}$.]
Thus for arbitrary field strengths, the condition $l_{\rm ph}<h$ 
leads to the constraint:
\be
\frac{\epsilon}{2m_ec^2}\beta_Q\,
\left(\frac{h}{{\cal R}_c}+\theta_e\right) \ga
{1\over 15}(1+15\beta_Q).
\label{chieq}
\ee

\subsection{Photon Emission Multiplicity and the Pulsar Death Line/Boundary}

There several possible photon emission mechanisms operating in the
vacuum breakdown, each leading to a different ``death line'', or more
precisely, ``death boundary''. We consider them separately.

\subsubsection{Curvature Radiation (CR)}

The characteristic energy of a photon emitted through curvature
radiation is $\epsilon \sim (3/2)\gamma^3 \hbar c/{\cal R}_c =
4.74\times 10^9 \beta_Q^3 h_3^6 P_0^{-3} {\cal R}_6^{-1}$~eV, where
${\cal R}_6={\cal R}_c/(10^6~{\rm cm})$, and we have used
$\gamma\sim\gamma_m$ [Eq.~(\ref{gammaeq})].  The emission angle is
$\theta_e\sim \gamma^{-1}$, which is typically much less than $h/{\cal
R}_c$ (this can be easily checked {\it a
posteriori}). Equation~(\ref{chieq}) then reduces to
\be
h > h_{\rm min,ph} = 546 P_0^{3/7} {\cal R}_6^{2/7} 
\left(\frac{15\beta_Q+1}{\beta_Q^4}\right)^{1/7}\mbox{ cm.}
\label{hphCReq}
\ee

The rate of energy loss of an electron or positron emitting curvature
radiation is $P_{\rm CR} = 2e^2\gamma^4/(3c^3)(c^2/{\cal R}_c)^2$, thus the 
number of photons emitted through curvature radiation
by a single electron or positron across the gap is 
\be
N_{\rm ph} \simeq {P_{\rm CR}\over\epsilon}{h\over c}\simeq
\frac{4}{9}\frac{e^2}{\hbar c}\frac{\gamma h}{{\cal R}_c} 
= 17.6 \beta_Q h_3^3 P_0^{-1} {\cal R}_6^{-1} \,.
\ee
The condition $N_{\rm ph}>\lambda$
[Eq.~(\ref{secondceq})] then gives
\be
h\ga h_{\rm min,e} = 
384 \lambda^{1/3} \beta_Q^{-1/3} P_0^{1/3} {\cal R}_6^{1/3} \mbox{ cm.}
\label{heCReq}
\ee
Thus the minimum gap height required for vacuum breakdown is 
$h \simeq \max(h_{\rm min,ph},h_{\rm min,e})$.
Combining Eqs.~(\ref{hmaxeq}), (\ref{hphCReq}), and (\ref{heCReq}),
we have
\be
\max(h_{\rm min,ph},h_{\rm min,e})< h_{\rm max}.
\label{CRdeatheq}
\ee
This gives a necessary condition for pulsar emission
and defines the pulsar ``death line''. For all relevant parameter regimes,
$h_{\rm min,ph}>h_{\rm min,e}$, and 
Eq.~(\ref{CRdeatheq}) simply becomes 
$h_{\rm min,ph}< h_{\rm max}$. The critical pulsar spin period is then
\be
P_{\rm crit} = 1.64\, b_d^{1/13}\, B_{12}^{8/13} {\cal R}_6^{-4/13} 
(1+15\beta_Q)^{-2/13}\mbox{ s,}
\label{PcritRSeq}
\ee
where the dipole polar field is $B_{12} = 2.0 (P_0 \dot{P}_{15})^{1/2}$,
with $\dot{P}_{15} = \dot{P}/(10^{-15}~\rm s~s^{-1})$.
For $\beta_Q\la 1/15$ this is the same as the result of \citet{ruderman75}.

\begin{figure}
\includegraphics[width=6.5in]{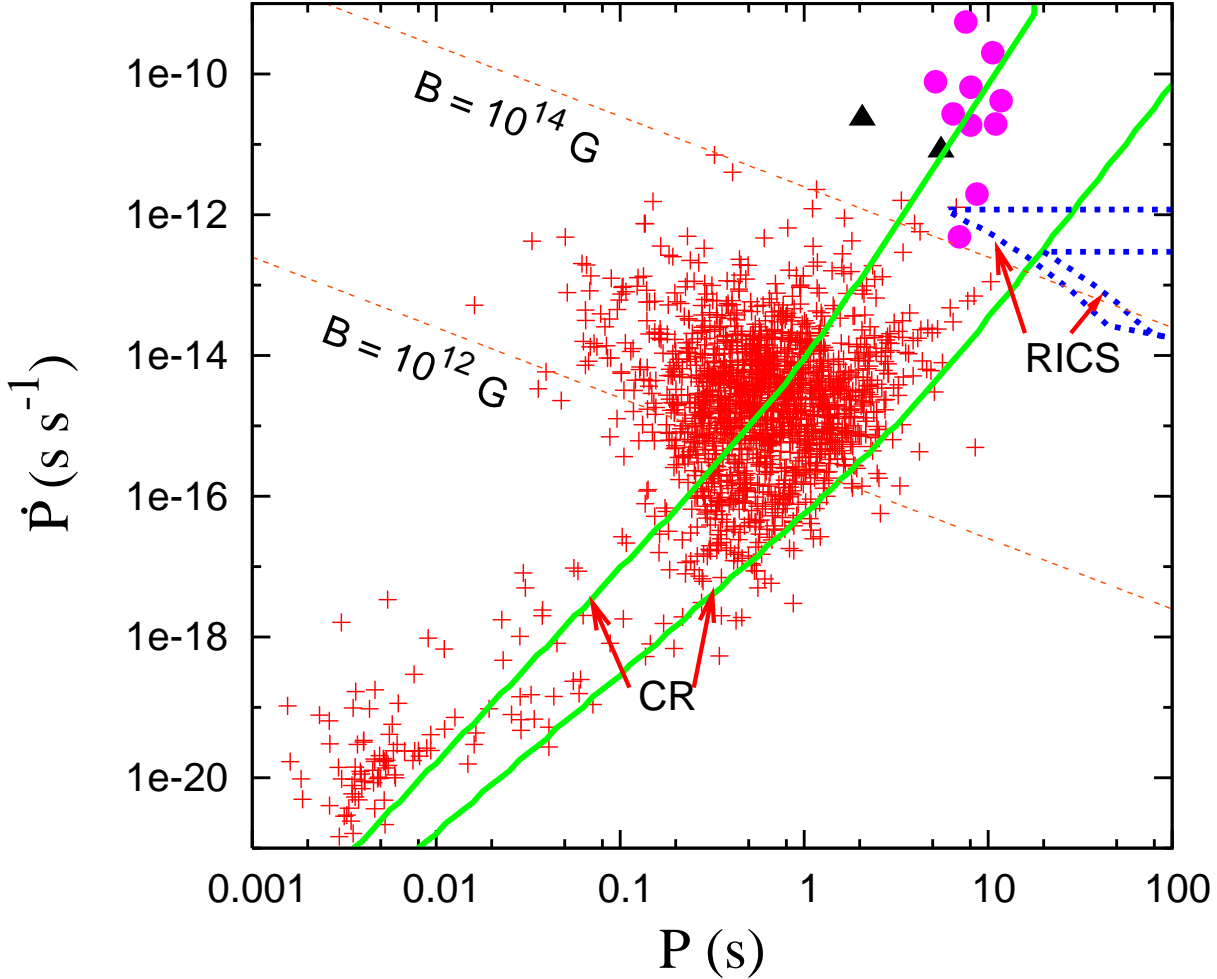}
\caption{Pulsar death lines/boundaries for the CR and resonant ICS gap
breakdown mechanisms. For curvature radiation, the lower line is for a
magnetic field radius of curvature comparable to the stellar radius
(${\cal R}_c\simeq R$) and the upper line is for a radius curvature
given by the dipole formula (${\cal R}_6=100 P_0^{1/2}$). For RICS,
the large ``box'' is for $\lambda=1$ and the small box is for
$\lambda=2$; both boxes are for a surface temperature of
$5\times10^6$~K. The unspecified neutron star parameters are taken to
be unity (i.e., we set $b_d=1$ and for RICS ${\cal R}_6=1$). The CR
mechanism operates (and the pulsar is alive) above and to the left of
the lines, and the RICS mechanism operates within the boxes.
Radio/X-ray pulsars (ATNF catalog,
http://www.atnf.csiro.au/research/pulsar/psrcat) are labeled by
crosses, while magnetars (McGill catalog,
http://www.physics.mcgill.ca/$\sim$pulsar/magnetar/main.html) are labeled
by solid circles and the two radio magnetars are labeled by solid
triangles.}
\label{deathfig}
\end{figure}

In Fig.~\ref{deathfig}, we show the death lines determined from
Eq.~(\ref{CRdeatheq}) for the cases of
${\cal R}_6=1$ and ${\cal R}_6=100P_0^{1/2}$ (pure dipole field at
the polar cap), with $b_d=1$.

\subsubsection{Resonant Inverse Compton Scattering (RICS)}
\label{sect:RICS}

Here the high-energy photons in the cascade are produced by Compton
upscatterings of thermal photons from the neutron star surface.
Resonant scattering in strong magnetic fields (e.g.,
\citealt{herold79}) can be thought of as resonant absorption (where
the electron makes a transition from the ground Landau level to the
first excited level) followed by radiative decay. Resonance occurs
when the photon energy in the electron rest frame satisfies
$\epsilon'\simeq \epsilon_c= \hbar (eB/m_ec)= \beta_Q m_ec^2$. The
resonant photon energy (in the ``lab'' frame) before scattering is
$\epsilon_{\rm i}=\epsilon_c/[\gamma(1-\cos\theta_i)]$, where
$\theta_i$ is the incident angle (the angle between the incident
photon momentum and the electron velocity). After absorbing a photon,
the electron Lorentz factor drops to
$\gamma_e=\gamma/(1+2\beta_Q)^{1/2}$, and then radiatively decays
isotropically in its rest frame. The characteristic photon energy
after resonant scattering is therefore (e.g., \citealt{beloborodov07})
\be
\epsilon = \gamma \left(1-\frac{1}{\sqrt{1+2\beta_Q}}\right) m_ec^2,
\ee
with typical emission angle $\theta_e\sim 1/\gamma_e$.
The condition $l_{\rm ph}<h$ [see Eq.~(\ref{chieq})] becomes 
\be
\frac{\gamma}{2}\left(1-\frac{1}{\sqrt{1+2\beta_Q}}\right)\beta_Q
\left(\frac{h}{{\cal R}_c}+\frac{\sqrt{1+2\beta_Q}}{\gamma}\right) \ga 
\beta_Q+\frac{1}{15}\,.
\label{betaRESeq}
\ee
For $\beta_Q \ga 4$ this condition is automatically satisfied, i.e.,
resonant ICS photons pair produce almost immediately upon being
upscattered. For $\beta_Q<4$, Eq.~(\ref{betaRESeq}) puts a constraint
on the gap height $h$. As we shall see below, most of the scatterings
in the gap are done by electrons/positrons with
$\gamma\sim\min(\gamma_c,\gamma_m)$, where $\gamma_c=\epsilon_c/kT$
(with $T$ the surface blackbody temperature) and $\gamma_m$ is the
Lorentz factor of a fully-accelerated electron or positron
[Eq.~(\ref{gammaeq})]. For $\gamma=\gamma_m$, Eq.~(\ref{betaRESeq})
yields
\be 
h \ga h_{\rm min,ph}^{(1)} = 56.9 P_0^{1/3} {\cal R}_6^{1/3}f(\beta_Q)^{1/3}~{\rm cm},
\label{eq:hminph1}
\ee
where
\be
f(\beta_Q)=
\frac{\sqrt{1+2\beta_Q}}{\beta_Q}\left(\frac{2}
{\sqrt{1+2\beta_Q}-1}{1+15\beta_Q\over 15\beta_Q}-1\right).
\label{eq:fbeta}\ee
For $\gamma=\gamma_c$ we have
\be 
h \ga h_{\rm min,ph}^{(2)} = 169 {\cal R}_6 T_6 f(\beta_Q)~{\rm cm}.
\label{eq:hminph2}
\ee
Combining Eqs.~(\ref{eq:hminph1}) and (\ref{eq:hminph2}), we find that
the condition $l_{\rm ph}<h$ leads to
\be
h \ga h_{\rm min,ph} = \max(h_{\rm min,ph}^{(1)},h_{\rm min,ph}^{(2)}) \,.
\label{eq:hminph}
\ee

The resonant cross section for inverse Compton scattering, in the rest
frame of the electron before scattering, is
\be
\sigma'_{\rm res}\simeq 2\pi^2 \frac{e^2 \hbar}{m_ec}\,
\delta(\epsilon'-\epsilon_c),
\ee
where $\epsilon'\sim\gamma\epsilon_i$.
This cross section is appropriate even for $B_p >B_Q$, since the
resonant condition $\epsilon' = \epsilon_c$ holds regardless of field strength
(cf.~\citealt{gonthier00}).
The ambient spectral photon number number density near the polar cap is
\be
{dn_{\rm ph}\over d\epsilon_i}= {\epsilon_i^2/(\pi^2\hbar^3c^3)\over 
e^{\epsilon_i/kT}-1}.
\ee
For concreteness, consider a positron produced at $z=0$ with initial
Lorentz factor $\gamma=1$ and accelerated to $\gamma=\gamma_m$ after
crossing the full gap.\footnote{We can also consider the general
situation where a positron (electron) is created at some location
within the gap with initial Lorentz factor much less than $\gamma_m$,
travels upwards (downwards) across the gap and get accelerated to a
final Lorentz factor of order $\gamma_m$. This would give similar
result for $N_{\rm ph}$.} Neglecting the radiation reaction (see
later), we have $\gamma-1=2(\gamma_m-1)(z/h-z^2/2h^2)$. The number of
photons upscattered through resonant ICS by the positron is given by
(see Appendix B)
\ba
N_{\rm ph} & \simeq & \int_0^h\! dz \, \int_0^\infty\! 
d\epsilon_i \,\frac{dn_{\rm ph}}{d\epsilon_i}\, \sigma'_{res} \nonumber\\
 & \simeq & {\beta_Q^2\over (\gamma_m-1)}{h\over a_0}
\int_1^{\gamma_m}\!{d\gamma\over \gamma^3\,\left(e^{\epsilon_c/kT\gamma}-1
\right)}\left(1-{\gamma-1\over \gamma_m-1}\right)^{-1/2}\nonumber \\
 & \simeq & \frac{1}{\gamma_m} \left(\frac{kT}{m_ec^2}\right)^2 \frac{h}{a_0}
\int_{x_m}^{\epsilon_c/(kT)}
\frac{x\,dx}{(e^x-1)(1-x_m/x)^{1/2}}
\label{eq:nph}\ea
where we have used $\gamma_m\gg 1$ and 
\be
x_m=\frac{\epsilon_c}{\gamma_m kT}=\frac{\gamma_c}{\gamma_m} = 1.09\times10^{-3} h_3^{-2} P_0 T_6^{-1}.
\label{xeq} 
\ee
Note that the second equality of Eq.~(\ref{eq:nph}) gives
\be
\frac{dN_{\rm ph}}{d\ln\gamma} \simeq \beta_Q^2 \frac{h}{\gamma_m a_0} \gamma^{-2} \left(e^{\gamma_c/\gamma}-1\right)^{-1}(1-\gamma/\gamma_m)^{-1/2} \,.
\label{eq:dNdlnG}
\ee
From this equation we see that for $\gamma_c = \epsilon_c/kT \la
\gamma_m$, $dN_{\rm ph}/d\ln\gamma$ peaks at $\gamma\sim\gamma_c$,
with $(dN_{\rm ph}/d\ln\gamma)_{\gamma=\gamma_c} \sim N_{\rm ph}$,
while for $\gamma_c<\gamma\la\gamma_m$, $dN_{\rm ph}/d\ln\gamma$ is of
order $(\gamma_c/\gamma)N_{\rm ph}$; for $\gamma_c\ga\gamma_m$,
$dN_{\rm ph}/d\ln\gamma \sim (\gamma/\gamma_m)N_{\rm ph}$ peaks at
$\gamma\sim\gamma_m$. Therefore, most of the scatterings in the gap
are done by electrons/positrons with
$\gamma\sim\min(\gamma_c,\gamma_m)$.
Since we are interested in the regime
$\epsilon_c/kT\gg 1$, the integral in the last equality of
Eq.~(\ref{eq:nph}) depends only on $x_m$, and for our purpose it can
be approximated as
$(\pi^2/6) x_m(e^{x_m}-1)^{-1}$. This approximation reproduces the
exact integral in the $x_m \rightarrow 0$ limit.
Thus we have 
\be 
N_{\rm ph,res} \simeq 4.89\times10^{-2}
\beta_Q^{-1} T_6^{5/2} P_0^{1/2} F(x_m)\,, \quad
{\rm with}~~~~F(x_m) =
\frac{x_m^{3/2}}{e^{x_m}-1}\,.  \ee 
The function $F(x_m)$ peaks at $x_m=0.874$ with
$F_{\rm max}=0.585$. Thus the condition $N_{\rm ph}>\lambda$ necessarily 
requires $2.86\times10^{-2} \beta_Q^{-1} T_6^{5/2} P_0^{1/2}\ga\lambda$, or 
\be
\beta_Q \la \beta_{Q,\rm crit} = 2.86\times10^{-2} \lambda^{-1}
T_6^{5/2} P_0^{1/2}\,.
\label{betacriteq}
\ee
For a given $\beta_Q<\beta_{Q,\rm crit}$, 
the condition $N_{\rm ph}>\lambda$ is equivalent to 
$F(x)> 0.588 \beta_Q/\beta_{Q,\rm crit}$, which limits $x_m$ to 
the range $x_a < x_m < x_b$, where $x_{a,b}$ are determining by
solving $F(x_m)= 0.588 \beta_Q/\beta_{Q,\rm crit}$. This condition
then translates to the constraint on $h$:
\be
h_{\rm min,e} < h < h_{\rm max,e} \,,
\ee
where
\be
h_{\rm min,e} = 33 x_b^{-1/2} P_0^{1/2} T_6^{-1/2}~{\rm cm}\,,\quad
h_{\rm max,e} = 33 x_a^{-1/2} P_0^{1/2} T_6^{-1/2}~{\rm cm}.
\label{eq:hminmax}\ee

In summary, vacuum breakdown involving RICS requires 
\be
\beta_Q<\beta_{Q,\rm crit}\quad{\rm and}\quad
\max(h_{\rm min,ph},h_{\rm min,e})<\min(h_{\rm max},h_{\rm max,e}),
\ee
where $\beta_{Q,\rm crit},~h_{\rm max},~
h_{\rm min,ph},~h_{\rm min,e},~h_{\rm max,e}$
are given by Eqs.~(\ref{betacriteq}), 
(\ref{hmaxeq}), (\ref{eq:hminph}) (note that $h_{\rm min,ph}=0$
for $\beta_Q\ga 4$), and (\ref{eq:hminmax}), respectively.
In Fig.~\ref{deathfig} we show the pulsar death boundary when RICS is
most important for initiating a cascade in the vaccum gap,
for the cases $\lambda=1$ and $\lambda=2$, with 
$b_d=1$, ${\cal R}_6=1$, and $T_6=5$.
Note that in Fig.~\ref{deathfig} we have not plotted RICS death
boundaries for the case of a dipole radius of curvature (${\cal R}_6=
100 P_0^{1/2}$) or a surface temperature $T_6 \la 1$; there are no
regions of the $P$--$\dot{P}$ diagram where vacuum gap pair cascades
are possible under these conditions.

The pulsar death boundary depicted in Fig.~\ref{deathfig} can be
understood as follows: (i) a) The condition
$h_{\rm min,ph}^{(1)}<h_{\rm max}$ gives
\be
{\rm (Ia)}\qquad P\la 352\,b_d^{-3/5}{\cal R}_6^{-2/5}{f(\beta_Q)}^{-2/5}~{\rm s},
\label{minphmaxeq}
\ee
where $f(\beta_Q)$ is given by Eq.~(\ref{eq:fbeta}). This is 
shown as the long-dashed line labeled (Ia) in Fig.~\ref{RICSfig}.
b) The condition $h_{\rm min,ph}^{(2)}<h_{\rm max}$ gives
\be
{\rm (Ib)}\qquad P\la 1.99\times10^3\,b_d^{-1}{\cal R}_6^{-2}T_6^{-2}{f(\beta_Q)}^{-2}~{\rm s}.
\label{minph2maxeq}
\ee
This is shown as the short-dashed line labeled (Ib) in Fig.~\ref{RICSfig}.
This set of conditions, (Ia) and (Ib), is the usual
requirement that photons emitted by an accelerated electron or positron
in the gap must be able initiate pair production.
(ii) a) For $\beta_Q\ll \beta_{Q,\rm crit}$, we have $x_a\simeq 
0.342\,(\beta_Q/\beta_{Q,\rm crit})^2$, and the condition 
$h_{\rm min,ph}^{(1)}<h_{\rm max,e}$ then yields
\be
{\rm (IIa)}\qquad P\ga 210\,\lambda^{3/2}{\cal R}_6^{1/2}T_6^{-3}\beta_Q^{3/2}f(\beta_Q)^{1/2}~{\rm s}.
\ee
This is shown as the dotted line labeled (IIa) in
Fig.~\ref{RICSfig}.
b) The condition $h_{\rm min,ph}^{(2)}<h_{\rm max,e}$ yields
\be
{\rm (IIb)}\qquad P\ga 105\,\lambda{\cal R}_6T_6^{-1}f(\beta_Q)~{\rm s}.
\ee
This is shown as the dot-long-dashed line labeled (IIb) in
Fig.~\ref{RICSfig}. This set of conditions, (IIa) and (IIb), together
with $\beta_Q\la \beta_{Q,\rm crit}$, come from the requirement for
efficient photon emission by RICS in the gap. (iii) The condition
$h_{\rm min,e}<h_{\rm max}$ gives
\be
{\rm (III)}\qquad P\la 228\,b_d^{-1/2}T_6^{1/2}x_b^{1/2}~{\rm s},
\quad {\rm with}~~x_b\sim 0.874+\ln {\beta_{Q,\rm crit}\over\beta_Q}.
\label{minemaxeq}
\ee
This condition is shown as the dot-short-dashed line labeled (III)
in Fig.~\ref{RICSfig}.
(iv) The condition $\beta_Q> \beta_{Q, \rm crit}$ gives Eq.~(\ref{betacriteq})
and is shown as the light solid line labeled (IV) in Fig.~\ref{RICSfig}.

Previous studies of the the pulsar death conditions for vacuum gaps
where RICS is the dominant photon emission mechanism have found that
the RICS mechanism can lead to gap breakdown for a wide range of
neutron star parameters (see, e.g., \citealt{zhang00}). This is
contrary to our results, which show (see Figs.~{\ref{deathfig} and
{\ref{RICSfig}) that RICS is not a good mechanism for gap breakdown,
except under very specific conditions (e.g., high surface temperatures
and long rotation periods). The discrepancy arises because previous
works did not calculate/estimate $N_{\rm ph}$ (the number of high energy
photons produced as a positron/electron crosses the gap)
correctly. For example, it was implicitly assumed that photon
production continues across the entire gap at the same rate as it does
when $\gamma\simeq\gamma_c$ (i.e., at the point of maximum RICS power
loss) \citep{zhang00}. This assumption is invalid for
$\gamma>\gamma_c$, as is discussed above: $dN_{\rm ph}/d\ln\gamma$ grows
with increasing gamma factor until $\gamma\sim\gamma_c$, and then it
decreases [see Eq.~(\ref{eq:dNdlnG})]; therefore, $dN_{\rm ph}/d\gamma$
(which is directly related to the photon production rate
$\dot{N}_{\rm ph}$) drops faster than $\gamma^{-1}$ above $\gamma\sim\gamma_c$.

Note that the accelerating positron/electron is not radiation-reaction
limited at $\gamma\simeq\gamma_c$, since the power loss due to RICS is
significantly smaller than the power gain due to traversal across the
potential drop. The power loss due to RICS is given by
\ba
P_{\rm loss} & = & c \int_0^\infty\! d\epsilon_i \,\frac{dn_{\rm ph}}{d\epsilon_i}\, \sigma'_{res} (\epsilon-\epsilon_i) \\
 & \simeq & \frac{2\beta_Q^2 c}{a_0}\left(1-\frac{1}{\sqrt{1+2\beta_Q}}\right) \frac{m_ec^2}{\gamma^2 \left(e^{\epsilon_c/kT\gamma}-1\right)} \,.
\ea
At the point of maximum RICS power loss
(when $\gamma=\gamma_c=\epsilon_c/kT$)
\ba
P_{\rm loss}(\gamma=\gamma_c) & \simeq & \frac{2c}{a_0}\left(1-\frac{1}{\sqrt{1+2\beta_Q}}\right) \left(\frac{kT}{mc^2}\right)^2 (e-1)^{-1} \,m_ec^2 \\
 & \simeq & 1.9\times10^{11} \left(1-\frac{1}{\sqrt{1+2\beta_Q}}\right) T_6^2 \,m_ec^2~{\rm s}^{-1}
\ea
(cf.~\cite{dermer90}). The power gain due to acceleration across the
gap is given by
\be
P_{\rm gain} = e{\cal E}_{\parallel}c = \frac{2\Omega\beta_Q}{\alpha a_0}(h-z) \,m_ec^2 \,.
\ee
Thus
\be
\left. \frac{P_{\rm gain}}{P_{\rm loss}} \right|_{\gamma=\gamma_c} \simeq 170 \left(\frac{h-z}{100~\rm cm}\right) P_0^{-1} T_6^{-2} \beta_Q \left(1-\frac{1}{\sqrt{1+2\beta_Q}}\right)^{-1} \,.
\ee
For most pulsar parameters, $P_{\rm gain} \gg P_{\rm loss}$ [e.g.,
in order for $\gamma$ to reach $\gamma_c$ the gap height must be at
least $h = 33 P_0^{1/2} T_6^{-1/2}$~cm; see Eq.~(\ref{xeq}) with $x_m=1$].
Therefore, there is no reason why $\gamma$ should remain near
$\gamma_c$, the point of maximum RICS photon emission, as was assumed
in some earlier papers.

\begin{figure}
\includegraphics[width=6.5in]{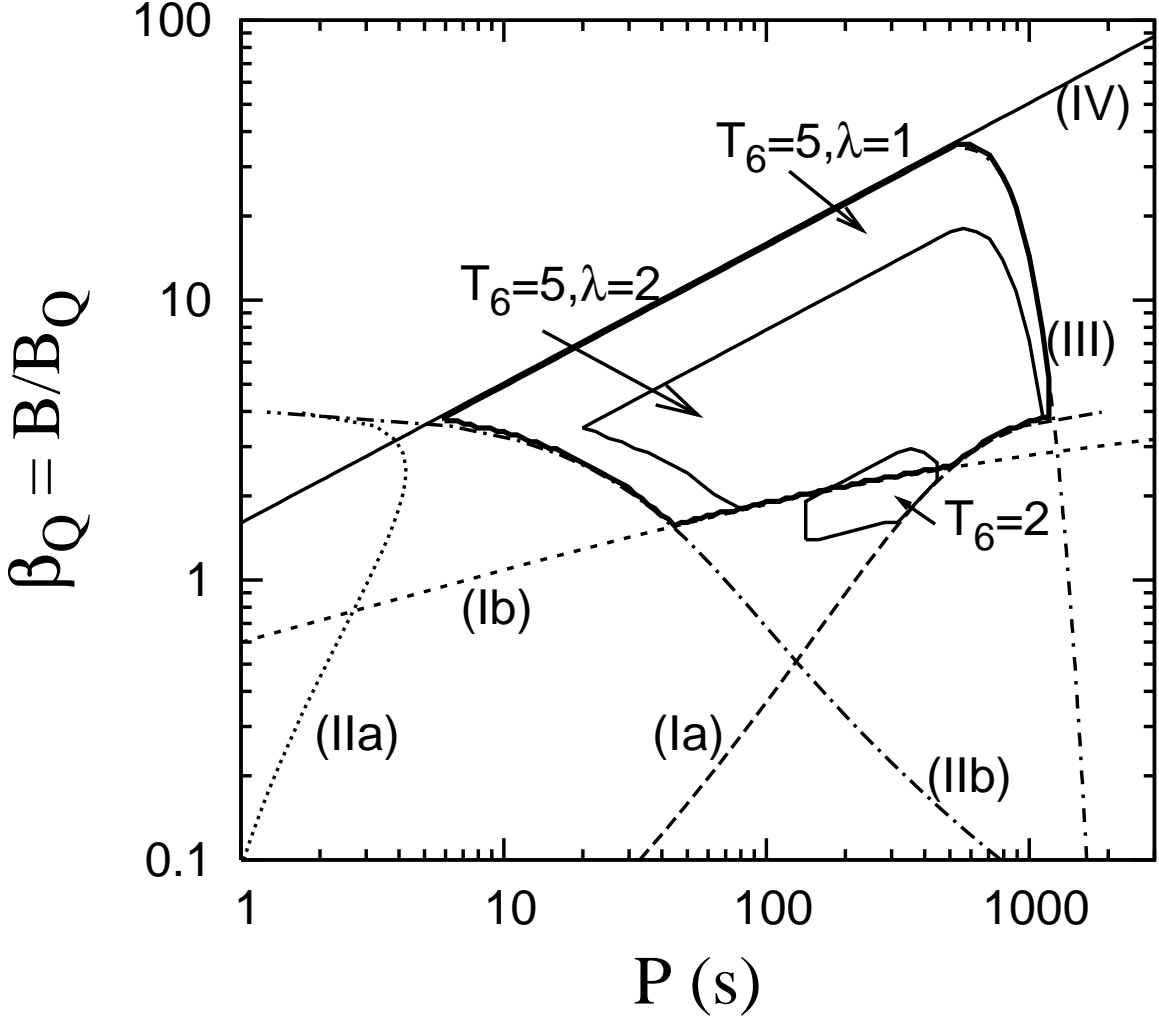}
\caption{The pulsar death boundaries when the resonant ICS mechanism
is most important for initiating a cascade, plotted as a function of
the dimensionless magnetic field strength $\beta_Q=B/B_Q$ and the
period $P$. The boundaries are shown for surface temperature
$T=5\times10^6$~K and parameter $\lambda=1$ (the largest, bold,
enclosed region), $T=5\times10^6$~K and $\lambda=2$ (the mid-sized
enclosed region) and $T=2\times10^6$~K and $\lambda=1$ (the smallest
enclosed region). The critical lines defining the edges of the region
for $T_6=5,\lambda=1$ are also shown. Each critical line (I)--(IV) is
determined by one of Eqs.~(\ref{minphmaxeq})--(\ref{minemaxeq}) and
(\ref{betacriteq}), as discussed in Section~\ref{sect:RICS}.}
\label{RICSfig}
\end{figure}

\subsubsection{Nonresonant Inverse Compton Scattering (NRICS)}
\label{sect:NRICS}

The characteristic energy of a photon Compton-upscattered by an
electron or positron of Lorenz factor $\gamma$ is $\epsilon\sim
\gamma\epsilon'/(1+x)$, where $x=\epsilon'/mc^2$, $\epsilon'\sim\gamma
\epsilon_i$, and $\epsilon_i$ is the initial seed photon energy; the
pitch angle of the scattered photon is of order
$\theta_e\sim(1+x)/\gamma$. In the vacuum gap, most the scatterings
are by electrons/positrons with $\gamma\sim\gamma_m$ on seed photons
with initial energy $\epsilon_i\sim 2.82kT$ (see below). Substituting
\be
\epsilon\sim {\gamma_m\epsilon'_m\over 1+x_m},
\qquad {\rm with}\quad x_m={\epsilon'_m\over mc^2}={2.82kT\gamma_m\over mc^2}
\label{eq:enric}\ee
into Eq.~(\ref{chieq}) (which results from the requirement $l_{\rm ph}
<h$), we find
\be
\frac{\gamma_m}{2}\left(\frac{x_m}{1+x_m}\right)\beta_Q
\left(\frac{h}{{\cal R}_c}+{1+x_m\over\gamma_m}\right) > \beta_Q+\frac{1}{15}.
\ee
Using Eq.~(\ref{gammaeq}), this becomes
\be
0.0415 \beta_Q^{-1/2} P_0^{1/2} T_6^{-3/2} {\cal R}_6^{-1} \frac{x_m^{5/2}}
{1+x_m} + x_m > 2\left(1+\frac{1}{15\beta_Q}\right)\,.
\label{FNReq}
\ee
The gap height is related to $x_m$ by 
\be
h = 19.7 x_m^{1/2} \beta_Q^{-1/2} P_0^{1/2} T_6^{-1/2}~{\rm cm.}
\label{hxNReq}
\ee
The solution to Eq.~(\ref{FNReq}) yields $x_m>x_{\rm min}$,
and thus the constraint on the gap height from $l_{\rm ph}<h$ is
\be
h \ga h_{\rm min,ph} = 
19.7 x_{\rm min}^{1/2} \beta_Q^{-1/2} P_0^{1/2} T_6^{-1/2}~{\rm cm.}
\ee

The nonresonant part of the ICS cross section, in the rest frame of
the electron before scattering, is approximately given by
\be
\sigma'(\epsilon')\simeq \sigma_T \left(\frac{\epsilon'}{\epsilon'+\epsilon_c}
\right)^2 f_{\rm KN}(x)
= \sigma_T \left(\frac{x}{x+\beta_Q}\right)^2 f_{\rm KN}(x) \,,
\ee
where $\sigma_T$ is the Thomson cross-section, $x=\epsilon'/(mc^2)$, and 
\be
f_{\rm KN}(x) =\frac{3}{4}\left[\frac{1+x}{x^3}\left\{\frac{2x(1+x)}{1+2x}
-\ln(1+2x)\right\}+\frac{1}{2x}\ln(1+2x)-\frac{1+3x}{(1+2x)^2}\right]
\ee
is the Klein-Nishina suppression factor [$f_{\rm KN}\simeq 1-2x$ for $x\ll 1$
and $f_{\rm KN}\simeq (3/8x)(\ln 2x+1/2)$ for $x\gg 1$].
This agrees well with the calculated NR cross sections in strong 
magnetic fields (e.g., \citealt{gonthier00}).

The number of scatterings per unit length by an electron or positron is
\be
{dN_{\rm ph}\over dz} \simeq \int_0^\infty d\epsilon_i \frac{dn_{\rm ph}}
{d\epsilon_i}\, \sigma'(\gamma\epsilon_i)\sim
0.24\left({kT\over \hbar c}\right)^3
\sigma'(2.82kT\gamma),
\label{npheq}
\ee
where in the second equality we have used the fact that $dn_{\rm
ph}/d\ln\epsilon_i$ peaks at $\epsilon_i=2.82kT$, while
$\sigma'(\epsilon')$ varies more slowly with $\epsilon'$. Similar to
Section~\ref{sect:RICS}, consider a positron produced at $z=0$ with
initial Lorentz factor $\gamma=1$ and accelerated to $\gamma=\gamma_m$
after crossing the full gap. The number of scatterings produced by the
positron is given by
\be
N_{\rm ph}\simeq \frac{h}{2\gamma_m} \int_1^{\gamma_m}
\frac{d\gamma}{\sqrt{1-\gamma/\gamma_m}} \,
{dN_{\rm ph}\over dz}.
\ee
Clearly, most of the scatterings are by positrons/electrons 
with $\gamma\sim \gamma_m$,
producing photons with energy $2.82kT\gamma_m^2/(1+x_m)$
[see Eq.~(\ref{eq:enric})]. The number of photons scattered by 
$\gamma=(0.7$--$1)\gamma_m$ electrons/positrons is
\ba
N_{\rm ph}&\sim& {h\over 2}\left({dN_{\rm ph}\over dz}\right)_{\gamma=\gamma_m}
\simeq 0.12\,h
\left({kT\over \hbar c}\right)^3\sigma_T
\left({x_m\over x_m+\beta_Q}\right)^2f_{\rm KN}(x_m)\nonumber\\
&\simeq& 1.3\times 10^{-4}
\beta_Q^{-1/2} T_6^{5/2} P_0^{1/2} F(x_m,\beta_Q),
\ea
where 
\be
F(x_m,\beta_Q)={x_m^{5/2}\over (x_m+\beta_Q)^2}\,f_{\rm KN}(x_m).
\ee

Now consider the vacuum breakdown condition $N_{\rm ph}>\lambda$.
For a given $\beta_Q$, the function $F(x_m,\beta_Q)$ has a maximum 
$F_{\rm max}(\beta_Q)$ (this maximum is approximately achieved at
$x_m\sim 2.24+3\beta_Q$).
Then $N_{\rm ph}>\lambda$ requires
\be
P\ga P_{\rm crit}(\beta_Q)
=5.7\times 10^7\lambda^2T_6^{-5}\beta_Q F_{\rm max}(\beta_Q)^{-2}~{\rm s}.
\ee
When this is satisfied, we additionally require
\be
{F(x_m,\beta_Q)\over F_{\rm max}(\beta_Q)}>\left[{P\over P_{\rm crit}(\beta_Q)}
\right]^{-1/2},
\ee
which yields the solution $x_a<x_m<x_b$. In terms of the gap height, we have
\be
h_{\rm min,e} < h < h_{\rm max,e} \,,
\ee
where
\be
h_{\rm min,e} = 19.7 x_a^{1/2} \beta_Q^{-1/2} P_0^{1/2} T_6^{-1/2}~{\rm cm,}
\quad 
h_{\rm max,e} = 19.7 x_b^{1/2} \beta_Q^{-1/2} P_0^{1/2} T_6^{-1/2}~{\rm cm.}
\ee
When the neutron star surface temperature $T_6 \le 5$ there are no
values of $\beta_Q$ or $P$ for which NRICS can initiate a cascade in
the vacuum gap. (Only when $T_6 \ga 9$ are there any $\beta_Q,P$ values
which permit an NRICS-initiated cascade, and even at these high
temperatures the allowed range of $\beta_Q$ and $P$ values is very small
and atypical of neutron stars.) Therefore, no pulsar death boundaries
appear for the NRICS process in Fig.~\ref{deathfig}.

\section{Discussion}
\label{sect:discuss}

It is well known that a strong magnetic field increases the binding
energy of individual atom and that of the zero-pressure condensed
matter. Very approximately, for $B\gg B_0$ [see Eq.~(1)], the former
increases as $(\ln B)^2$ while the latter scales as
$B^{0.4}$. Therefore one expects that the outermost layer of a neutron
star may be in the condensed state when the magnetic field $B$ is
sufficiently strong and/or the surface temperature $T$ is sufficiently
low. Exactly under what conditions this occurs is an important
question that entails quantitative calculations. In this paper, using
our recent results on the cohesive properties of magnetized condensed
matter \citep{medin06a,medin06b}, we have established quantitatively
the parameter regime (in $B$ and $T$ space) for which surface
condensation occurs. Our calculations showed that there are a range of
neutron star magnetic field strengths and surface temperatures where
the condensed surface will have an important effect on radiation from
these stars. For example, if the surface composition is Fe, then
strong-field neutron stars ($B\ga10^{13}$~G) with moderate
($T\la10^6$~K) surface temperatures should have atmospheres/vapors
that are effectively transparent to thermal radiation, so that the
emission becomes that from a bare condensed surface. This may explain
the nearly blackbody-like radiation spectrum observed from the nearby
isolated neutron star RX J1856.5-3754 (e.g.,
\citealt{burwitz03,vanadelsberg05,ho07}).

We have also examined the conditions for the formation of a vacuum
acceleration gap above the polar cap region of the neutron star. The
inner acceleration gap model, first developed by \citet{ruderman75},
has provided a useful framework to understand numerous observations of
radio pulsars. Most notably, the model naturally explains the
phenomenon of drifting subpulses observed in many pulsars (e.g.,
\citealt{backer76,deshpande99,weltevrede06}) in terms of the ${\bf
E}\times{\bf B}$ circulation of plasma filaments produced by vacuum
discharges. Partially screened gaps have also been studied (e.g.,
\citealt{cheng80,gil03,gil06}). However, it has long been recognized
that the original Ruderman \& Sutherland model is problematic since
the dipole magnetic field inferred from $P,\dot P$ may not be strong
enough to inhibit charge emission from the surface.  Our calculations
described in this paper quantify the condition for vacuum gap
formation (see Fig.~\ref{gapfig}). While this condition (i.e., $T$ is
smaller than a critical value which depends on $B$ and composition)
may not be satisfied for most pulsars (unless one invokes surface
magnetic fields much stronger than that inferred from $P,\dot P$; see
Gil et al.~2006 and references therein), it could well be satisfied
for some neutron stars. In particular, the recently discovered high-B
radio pulsars, having dipole surface magnetic fields in excess of
$10^{14}$~G and temperature about $10^6$~K (e.g.,
\citealt{kaspi04,mclaugh05}), may operate a vacuum gap accelerator. On
the other hand, while magnetars have similar magnetic field strengths,
their surface temperatures are about five times larger than those of
high-B radio pulsars, and therefore may not have a vacuum gap. In this
regard, it is interesting to note that most magnetars do not show
radio emission (though this may be because the radio pulse is beamed
away from us or the because their magnetosphere plasma ``overwhelms''
the radio pulses), and the two recently detected radio magnetars have
rather different radio emission properties (e.g., the spectrum extends
to high frequency and the radiation shows high degrees of linear
polarization) compared to ``normal'' radio pulsars. We may therefore
speculate that a key difference between magnetars and high-B radio
pulsars is their difference in surface temperature. In any case, our
gap formation condition (Fig.~\ref{gapfig}) suggests that the radio
emission property of neutron stars may depend not only on the magnetic
field and rotation rate, but also on the surface temperature.

We note that our calculation of the requirements for vacuum gap
formation assumes idealized conditions. A real neutron star polar cap
may be immersed in a strong radiation field and suffer bombardment
from high energy particles (e.g.,
\citealt{arons81,beloborodov07}). The effective cohesive energy of the
surface may be somewhat smaller than what we used in our paper due to
surface defects (Arons 2007, private communication). Whether the
vacuum gap survives in realistic situations is unclear. It has been
suggested that a partially screened gap is formed instead
\citep{gil03,gil06}. With small modifications [e.g., the potential
drop given by Eq.~(\ref{eq:Phidrop}) is reduced], our discussion of
pair cascades in the vacuum gap can be easily generalized to the case
of a partially screened gap.

A major part of our paper is devoted to the pair cascade physics in
the vacuum gap (Section~\ref{sect:death}). We find that pair cascade
initiated by curvature radiation can account for most pulsars in the
$P$--$\dot{P}$ diagram, but significant field line curvature near the
stellar surface is needed. Although such field curvature is possible
for some pulsars, it is unlikely to occur for all of them. For a pure
dipole magnetic field, only about half of all pulsars can be explained
by a curvature radiation-initiated cascade. Contrary to previous works
(e.g., \citealt{zhang00}), we find that inverse Compton scatterings
(resonant or not) are not efficient in producing vacuum breakdown via
pair cascade.

The recent detection of the radio emission from two AXPs
(\citealt{camilo06,camilo07}) is of great interest. The emission
appears to be triggered by X-ray outbursts of usually quiescent
magnetars. This may be due to a rearrangement of the surface magnetic
field, which made pair cascades possible. We note that the occurrence
of pair cascades depends strongly on the field line
geometry/curvature; our study of pair cascades in the context of
vacuum gap accelerators (Section~\ref{sect:death}) serves as an
illustration of this point.

\section*{Acknowledgments}

This work has been supported in part by NASA Grant NNX07AG81G, NSF
grants AST 0307252 and 0707628, and by {\it Chandra} grant TM6-7004X
(Smithsonian Astrophysical Observatory).

\appendix

\section{Maximum Potential Drop For an Oblique Rotator}

For an oblique rotator, with the magnetic inclination angle $\alpha$
much larger than the polar cap angular size $r_p/R$, the voltage drop
across the polar cap is of order $(\Omega B_p/c)R r_p\sin\alpha$,
which is a factor of $R/r_p$ larger than the aligned case. Here we
show explictly that the maximum potential drop across the height of
the vaccum gap is still of order $(\Omega B_p/c) r_p^2$.

We will be working in the ``lab'' frame, where the star is rotating.
For simplicity we approximate the vacuum gap to be a cylinder of radius
$r_p$ and height $h \ll R$; see Fig.~\ref{fig:polargap}. In reality
the bases of the cylinder are not exactly circular for an oblique rotator,
but this does not affect our conclusion. The gap is small compared to the
stellar radius and we can treat it locally, using a Cartesian coordinate
system: $z$ along the gap height and $x,y$ for the
distance from the pole (with $\hat{x}$ being principally along
$\hat{\theta}$ and $\hat{y}$ along $\hat{\phi}$; $\hat\theta$ points in the
direction from the rotational pole to the magnetic pole).
The magnetic field in the cylinder is approximately uniform, $\mathbf{B} =
-B_p\hat{z}$.

\begin{figure}
\includegraphics[width=6.5in]{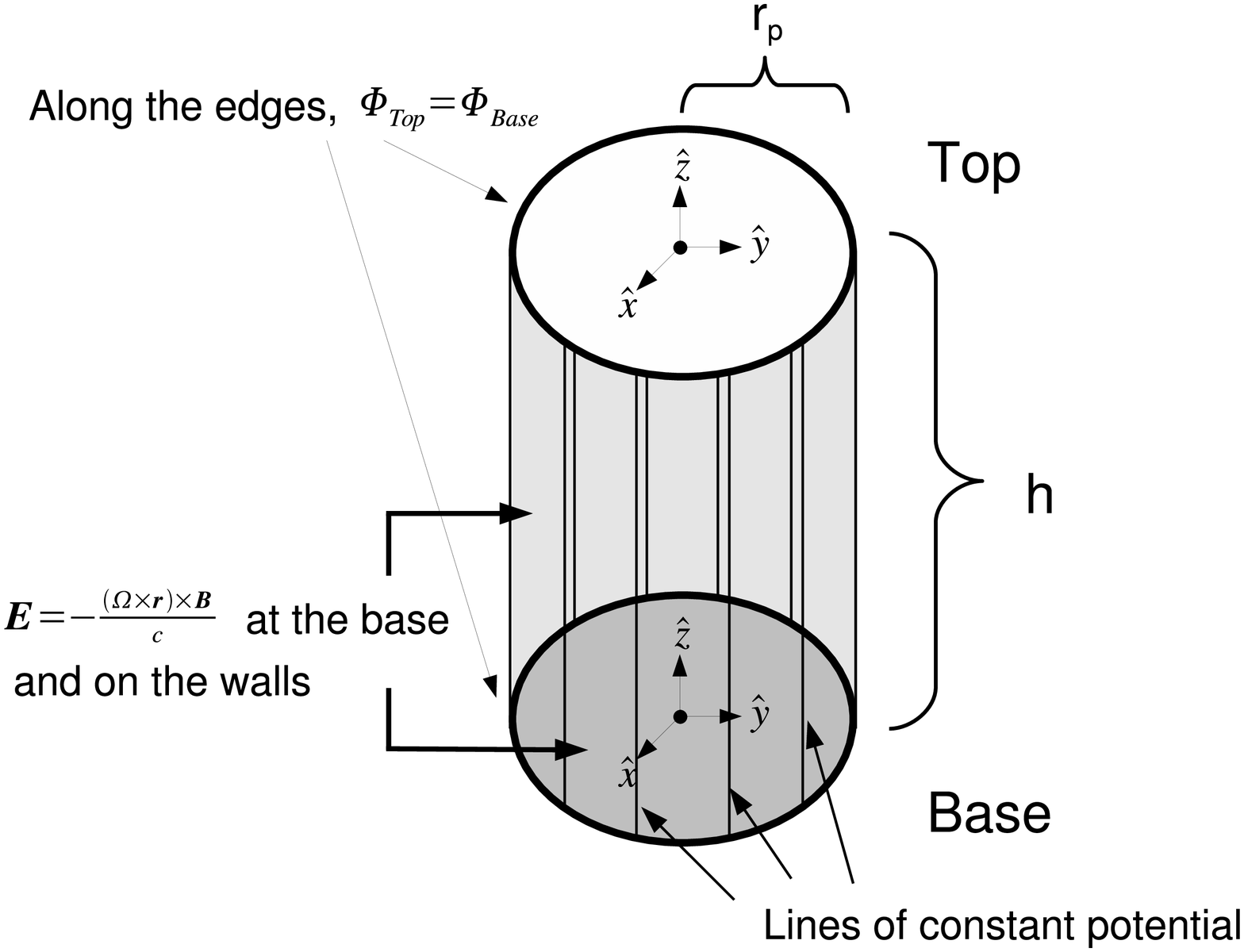}
\caption{A schematic diagram showing the polar gap structure in the
cylindrical approximation.}
\label{fig:polargap}
\end{figure}

The electric potential inside the cylindrical gap satisfies
$\nabla^2\Phi=0$.  The potential at the base and on the walls of the
cylinder can be found from $\mathbf{E}=-c^{-1}
[(\mathbf{\Omega}\times\mathbf{r})\times\mathbf{B}]$. At the top of
the cylinder, the potential satisfies $E_z=-\partial\Phi/\partial
z=0$.  With these boundary conditions the potential within the
cylinder is completely determined.

Without solving the complete potential problem, here we only consider
the potential at the top of the cylinder in the limit of $h \gg r_p$
(but $h \ll R$); this
corresponds to the maximum potential drop across the gap.
In this limit, the electric field on the top satisfies not only
$E_z=0$ but also $dE_z/dz=0$. Thus the Laplace equation below the top
of the cylinder becomes
\be
\frac{d^2\Phi}{dx^2} + \frac{d^2\Phi}{dy^2} = 0 \,.
\label{eq:Laplace}
\ee
To lowest order in $x/R$ and $y/R$ the electric field below the base
of the cylinder is given by
\be
\mathbf{E} = -\frac{(\mathbf{\Omega}\times\mathbf{r})\times\mathbf{B}}{c}
\simeq \frac{\Omega B_p R}{c} \left[\left(\sin\alpha+\cos\alpha\frac{x}{R}\right)\hat{x} + \cos\alpha\frac{y}{R}\,\hat{y}\right] \,.
\label{eq:approxE}
\ee
The potential at the base of the cylinder is therefore (using
${\bf E} = -\nabla\Phi$ and renormalizing such that the potential is
zero at the pole)
\be
\Phi_{\rm base} = -\frac{\Omega B_p R^2}{c} \left[\sin\alpha\frac{x}{R} + \cos\alpha\left(\frac{x^2}{2R^2}+\frac{y^2}{2R^2}\right)\right] \,.
\label{eq:baseV}
\ee
Since $E_z = 0$ on the cylindrical wall,
the potential on the wall is also given by Eq.~(\ref{eq:baseV}).
The potential at the top of the cylinder must
solve Eq.~(\ref{eq:Laplace}) and match the potential on the wall
along the upper edge. For a circular polar cap boundary,
given by $x^2+y^2=r_p^2$, the potential at the top is then
\be
\Phi_{\rm top} = -\frac{\Omega B_p R^2}{c} \left[\sin\alpha\frac{x}{R}+\cos\alpha\frac{r_p^2}{2R^2}\right] \,.
\label{eq:topV}
\ee


\begin{figure}
\includegraphics[width=5in]{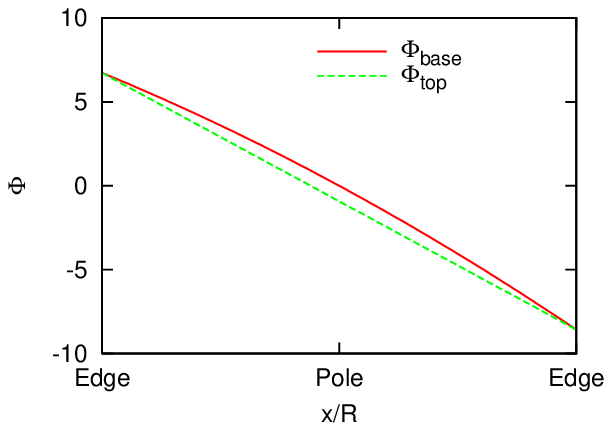}
\caption{Potential along the $\hat{x}$ ($\sim\hat{\theta}$) direction,
through the magnetic pole, both at the stellar surface and at the top of the
vacuum gap, for an oblique rotator. The magnetic inclination angle is chosen
to be $\alpha=\pi/8$. The potential is measured in units of
$\Omega B_p r_p^2/2c$, the value of the maximum potential drop for an
aligned rotator case.}
\label{fig:Vx}
\end{figure}


From Eqs.~(\ref{eq:baseV}) and (\ref{eq:topV}), we find that at the
magnetic pole, $|\Phi_{\rm top}-\Phi_{\rm base}| \simeq
(\Omega B_p r_p^2/2c) \cos\alpha$, which is the value of the
aligned case multiplied by $\cos\alpha$. Figure~\ref{fig:Vx}
compares the potential at the base and top of the
cylindrical gap along the $x$ axis. We see that although there is
a large potential drop across the polar cap, the potential difference
between the top and the base is smaller.

Alternatively, we may examine the problem in the rotating frame, in
which the potential inside the vacuum gap satisfies the equation
$\nabla^2\Phi=4\pi\rho_{\rm GJ}$, where $\rho_{\rm GJ}$ is the
Goldreich-Julian charge density. The potential at the base and on the
wall of the cylinder is $\Phi=0$ (since the electric field is zero
there). At the top of the cylinder, we have $\partial\Phi/\partial
z=0$. These boundary conditions completely determine the potential
inside the cylinder. For $h\ll r_p$, we expect the potential drop
along the $z$-axis, $\Delta\Phi$, to grow as $h^2$. But when $h$
becomes larger than $r_p$, the potential drop $\Delta\Phi$ will
saturate to $(\Omega B_p r_p^2/2c) \cos\alpha$, similar to the aligned
case.

\section{Scattering Rate Calculation}

In the neutron star rest frame (``lab'' frame), the electron
(positron) is embedded in a radiation field with specific intensity
$I_{\epsilon_i}(\hat{\Omega}_i)$. In the electron rest frame, the
radiation intensity is
\be
I'_{\epsilon'}(\hat{\Omega}') = \left(\frac{\epsilon'}{\epsilon_i}\right)^3 I_{\epsilon_i}(\hat{\Omega}_i) \,,
\label{eq:intensity}
\ee
where $\epsilon'$ and $\epsilon_i$ are related by a Lorentz
transformation: $\epsilon' = \epsilon_i \gamma(1-\beta\cos\theta_i)$.
For a photon coming in along the $\hat{\Omega}'$ direction, the
total scattering cross section is
$\sigma' = \int d\Omega'_1 \left(\frac{d\sigma}{d\Omega'_1}\right)_{\hat{\Omega}'\rightarrow\hat{\Omega}'_1}$,
which in general depends on $\hat{\Omega}'$ and $\epsilon'$. The
scattering rate in the electron rest frame is
\be
\frac{dN}{dt'} = \int d\Omega' \int d\epsilon' \,\frac{I'_{\epsilon'}}{\epsilon'} \,\sigma' \,.
\ee
In the lab frame the scattering rate is
$dN/dt = \gamma^{-1} (dN/dt')$ (e.g., \citealt{rybicki79}). Using
$d\Omega'/d\Omega_i = (\epsilon_i/\epsilon')^2$ and Eq.~(\ref{eq:intensity})
we have
\be
\frac{dN}{dt} = \int d\Omega_i \int d\epsilon_i \,(1-\beta\cos\theta_i) \,\frac{I_{\epsilon_i}}{\epsilon_i} \,\sigma' \,.
\ee
Neglecting the angle dependence of $\sigma'$ and assuming that the
radiation field $I_{\epsilon_i}$ is isotropic, this becomes
\be
\frac{dN}{dt} \simeq c \int d\epsilon_i \frac{dn_{\rm ph}}{d\epsilon_i} \sigma' \,,
\ee
which is the same as Eq.~(\ref{eq:nph}).

\label{lastpage}

\end{document}